# Nanoscale Fluorescence Thermometry: Probes, Recent Advances and Emerging Directions

*Md Shakhawath Hossain,[1] Nhat Minh Nguyen,[1] Thi Ngoc Anh Mai,[1] Trung Vuong Doan,[1] Chaohao Chen,[2] Qian Peter Su,[2] Jiayan Liao,[3] Yongliang Chen,[4] Quynh Le-Van,[5] Vu Khac Dat,[5] Toan Dinh,[6,7] Xiaoxue Xu,[2] and Toan Trong Tran[1,*]*

[1]School of Electrical and Data Engineering, University of Technology Sydney, Ultimo, NSW, 2007, Australia.

[2]School of Biomedical Engineering, University of Technology Sydney, Ultimo, NSW, 2007, Australia.

[3]School of Mathematical and Physical Sciences, Faculty of Science, University of Technology Sydney, NSW 2007, Australia

[4]Department of Physics, The University of Hong Kong, Pokfulam, Hong Kong, China.

[5]Center for Materials Innovation and Technology (CMIT), College of Engineering and Computer Science, VinUniversity, Gia Lam District, Hanoi 10000, Vietnam

[6]School of Engineering, University of Southern Queensland, Toowoomba, Queensland 4350, Australia

[7]Center for Future Materials, University of Southern Queensland, Toowoomba, Queensland 4350, Australia

*Corresponding author: trongtoan.tran@uts.edu.au

**Funding:** T. T. T acknowledges the financial support from the Australian Research Council (DE220100487, DP240103127). T. T. T. and T. D. thank the Queensland Department of Environment, Science, and Innovation for their financial support (Q2032010). This project was funded by the Queensland Government through the Department of Environment, Tourism, Science and Innovation's (DETSI) Quantum 2032 Challenge Program. The program aims to accelerate the development of quantum-based innovations in sports tech and related fields, foster collaboration between Queensland's quantum research sector and industry, and showcase the state's quantum expertise on the global stage during the Brisbane 2032 Olympic and Paralympic Games, contributing to the lasting legacy of the Games. This research is supported by an Australian Government Research Training Program (RTP) Scholarship.

## Abstract

The transition of materials and devices to nanometer, atomic, and quantum scales makes thermal characterization increasingly challenging, driving the need for advanced nanoscale thermometry. Fluorescence nanothermometry has emerged as a powerful approach, enabling remote, spatially resolved temperature measurements with sub-micrometer-to-nanometer precision across applications in nanoelectronics, microfluidics, and biological systems. In these systems, temperature is inferred from variations in fluorescence observables, including spectral position, intensity, linewidth, and excited-state dynamics. This review provides a comprehensive and critical overview of fluorescence nanothermometry, covering fundamental mechanisms, material platforms, recent advances, and emerging applications. It further presents a critical evaluation of key challenges and discusses emerging strategies and future research directions toward achieving robust, real-time thermometry. It is anticipated that this review will stimulate further advances in material platforms and system design, accelerating the development of accurate, scalable, and application-ready nanoscale thermometers.

## 1. Introduction

Temperature, classically described as the average kinetic energy of the constituent particles in a system, is one of the most frequently measured and fundamentally important physical quantities across science, health, and technology [1-4]. Precise temperature measurement, commonly known as thermometry, is essential across many sectors, including manufacturing, energy systems, aerospace, nuclear engineering, climate monitoring, cell biology, and healthcare [4-8]. Temperature is conventionally measured using a contact-based probe that physically contacts the object or body whose temperature is unknown. According to Mordor Intelligence, the global temperature sensor market, which is moderately concentrated, is valued at USD 9.93 billion in 2026 and is projected to reach USD 13.41 billion by 2031, with a compound annual growth rate (CAGR) of 6.2% [9]. Contact-based technologies such as thermocouples, thermistors, and resistance temperature detectors (RTDs) continue to dominate this sector, accounting for approximately 39.9% of total revenue in 2025 [9]. However, because contact-based methods rely on physical contact with the sample, they become impractical at micro and nanoscales, where the probe can introduce thermal loading and distort the true temperature distribution. Additionally, the need for an electrical link between the probe and the readout system limits their use on moving objects and in electrically noisy or spark-sensitive environments [10]. Moreover, at micro- and nanometer length scales, heat transport deviates fundamentally from the diffusive behavior that dominates at the macroscopic level. When the characteristic dimension of a heat source becomes comparable to or smaller than the mean free path (MFP) of the heat-carrying particles, and when temperature gradients exist over distances shorter than these MFPs, heat carriers experience substantially fewer scattering events. This results in a transition from diffusive to quasi-ballistic heat transport. Under such conditions, Fourier's law, which assumes local thermal equilibrium and diffusive transport, fails to accurately describe the heat dissipation process. Instead, a rigorous treatment based on the Boltzmann transport equation (BTE) is required to capture the non-diffusive nature of heat flow [11]. This introduces significant challenges in quantitatively reconciling experimental observations with theoretical predictions.

As a result, new strategies are needed to measure and manage heat in nanoscale systems, particularly those that rely on non-contact and remotely operated detection methods. Nanoscale thermometry has emerged to probe and understand thermodynamic mechanisms that conventional macroscale approaches cannot adequately describe. Among the various nanoscale temperature-sensing modalities, fluorescence-based thermometry has attracted particular attention due to its high sensitivity, compatibility with optical microscopy, and the ability to enable remote optical temperature readout, even when the sensing probe is in physical contact with the sample.

Fluorescence-based nanoscale thermometry has been the focus of several review articles over the past few years, providing broad overviews of materials platforms and representative applications [12-23]. Existing reviews on fluorescence-based nanoscale thermometry largely concentrate on biological and biomedical applications, with particular emphasis on luminescence thermometry based on transition-metal (TM) and trivalent lanthanide ($Ln^{3+}$) ions. In this context, this review begins with a comparative overview of the main micro/nanothermometry strategies, establishing a unified framework for assessing their operating principles, strengths, and limitations. Building on this foundation, the review is primarily dedicated to fluorescence-based thermometry, with a specific focus on diamond color-center thermometers, which represent some of the most advanced and versatile platforms for micro- and nanoscale temperature sensing. To ensure completeness and balance within the broader fluorescence thermometry landscape, alternative strategies based on quantum dots (QDs) and upconversion nanoparticles (UCNPs) are also discussed. Through a critical discussion of emerging trends, new possibilities, and representative applications, this review aims to present an updated perspective on fluorescence thermometry and to serve as a useful reference for students, researchers, and practitioners, while encouraging continued progress in the field.

## 2. Thermometry Techniques from Conventional to Micro/Nanoscale: Principles and Comparisons

Since Galileo's invention of the thermoscope, temperature measurement techniques have evolved significantly, leading to the development of a wide range of thermometry methods and sensors [24]. These methods can be classified according to the underlying physical mechanism, the nature of the detected signal, or whether the measurement requires physical contact with the sample [25].

### 2.1. Classification with Underlying Physical Mechanism

With the underlying physical mechanism, thermometry can be classified into primary and secondary thermometers. A primary thermometer does not require calibration because it directly measures temperature in accordance with a fundamental physical law. Examples include the constant-volume gas thermometer [26] (based on the ideal gas law), the Johnson-noise thermometer [27] (which extracts temperature from thermally induced voltage fluctuations in a resistor), the shot-noise thermometer [28] (which relates temperature to the bias voltage by measuring the electrical noise generated in a tunnel junction.), the acoustic thermometer [29] (based on the temperature dependence of the speed of sound in gas), and Coulomb-blockade thermometry [30] (which determines temperature from the charging effects in nanoscale devices containing multiple tunnel junctions). In each case, the thermometer uses a distinct physical law with a direct analytical form $T = \phi(q_1, q_2, \ldots)$. Although primarily used in metrology, primary thermometers play a crucial role in the ongoing redefinition of the International Temperature Scale (ITS-90) based on the Boltzmann constant ($\boldsymbol{k_B}$) [31]. However, the practical limitations of primary thermometers (cost, reproducibility, speed, and ease of use) make it essential to employ secondary thermometers in most applications [32]. Unlike primary thermometers, secondary thermometers cannot determine temperature directly because their measured properties do not follow the universal temperature law. Instead, they sense a temperature-dependent signal that must be calibrated to obtain the actual temperature. This calibration is performed using fixed, well-defined reference points standardized in ITS-90 [33]. Examples of secondary thermometers include thermocouples, RTDs, thermistors, liquid-in-glass thermometers, luminescence-based thermometers, etc. Even though secondary thermometers are less complex than primary ones, they require repeated calibration when used in different environments. Such recalibration is often impractical for luminescent thermometers in living cells or operating electronic devices, leading many studies to assume a single, medium-independent calibration relation, which may introduce inaccuracies [12].

### 2.2. Contact and Non-Contact Measurement Techniques

Thermometry can also be classified into contact and non-contact methods, depending on whether the sensor physically touches the sample [34]. Contact-based thermometers include electrical sensors such as resistance temperature detectors, thermistors, and thermocouples, as well as scanning probe methods such as scanning thermal microscopy (SThM). Contact-based thermometers suffer from thermal loading that can disturb the local temperature field, limited spatial resolution due to their finite dimensions, and incompatibility with sensitive or confined environments such as biological systems, thereby driving the need for optical, non-invasive thermometry approaches. Non-contact optical thermal sensors, which encompass fluorescence, Raman, Infrared (IR), and fiber-optic-based sensing approaches, are also expanding rapidly and are forecast to reach USD 3.5 billion by 2033, with a 7.8% CAGR, according to Strategic Revenue Insights [35]. Although often described as non-contact, fluorescence thermometry is not entirely contact-free because the fluorescent material must be placed directly on the sample, making it semi-contact. However, the temperature readout itself remains non-contact, as the detection system interacts only optically with the luminescent probe and has no physical connection to the sample. Conversely, IR thermometry, Raman-based thermometry, and emerging transducer-less thermoreflectance techniques can be regarded as truly non-contact methods.

### 2.3. Classification with Nature of the Signal

Another way to classify thermometry is by the nature of the signal used for temperature sensing. Broadly, for nanoscale applications, these techniques fall into two main categories: optical and non-optical systems [13]. This distinction is particularly important for certain applications, as all-optical methods are generally noninvasive and enable remote temperature readout, making them well-suited for cell-level temperature monitoring. In contrast to optical thermometry, non-optical methods are typically invasive, such as thermocouples and scanning thermal microscopy, which require the temperature probe to be in direct physical contact with the sample.

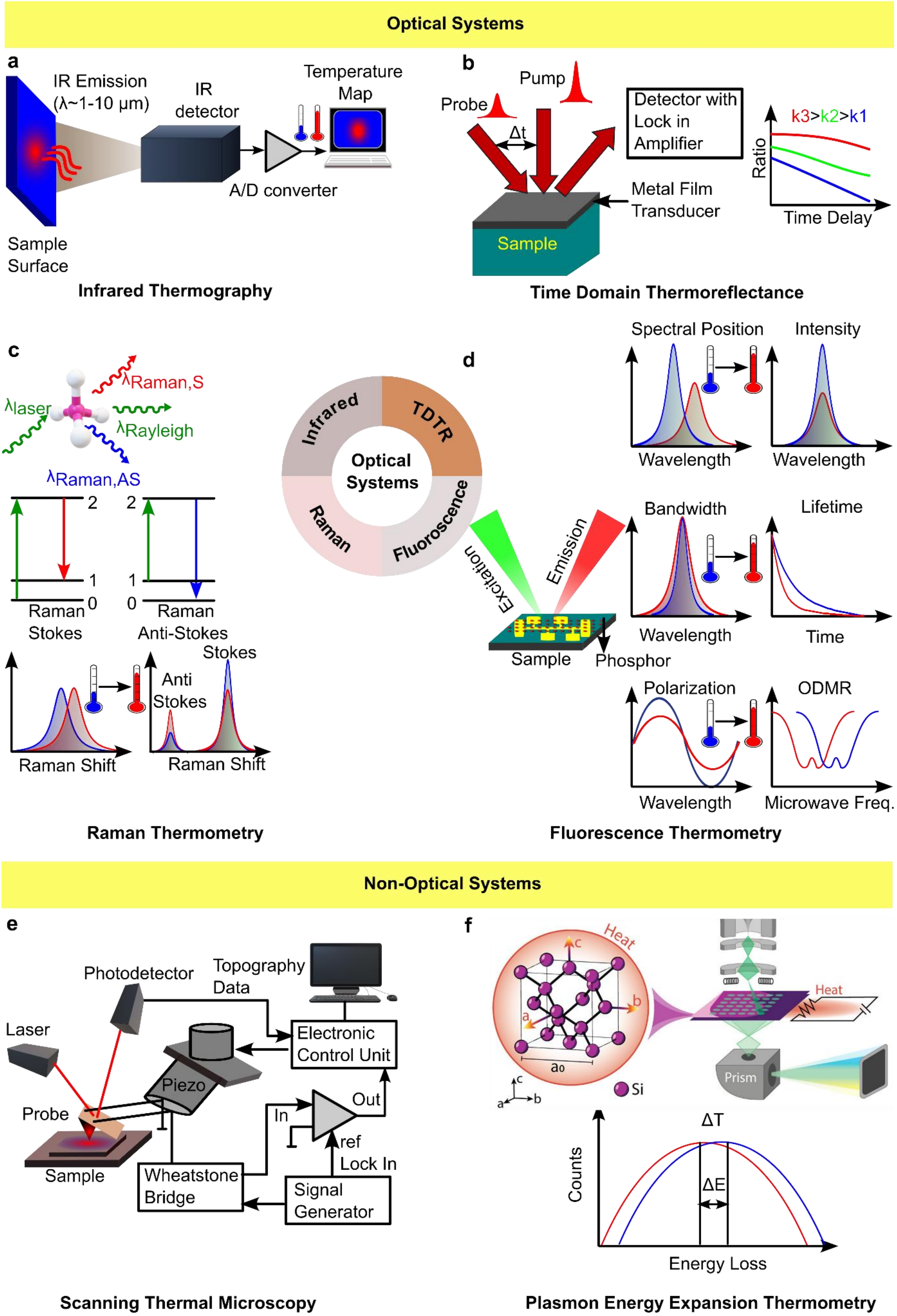

**Figure 1.** Overview of micro/nanoscale thermometry strategies, broadly classified into optical and non-optical approaches. **(a)** Schematic of the basic principle of Infrared thermography, illustrating temperature mapping based on thermal radiation emitted from the sample. **(b)** Schematic of time-domain thermoreflectance, showing ultrafast pump–probe reflectivity measurements used to extract thermal conductivity and heat capacity of thin films and bulk samples. (**c)** Raman thermometry principle, where temperature is extracted from temperature-dependent Raman peak shifts and intensity ratios. **(d)** Fluorescence thermometry principle, based on temperature-dependent changes in fluorescence spectral features. (**e)** Schematic illustration of scanning thermal microscopy setup. (**f)** Plasmon energy expansion thermometry: Measurement of resistive-heating-induced thermal expansion in crystalline silicon using scanning TEM-electron energy-loss spectroscopy with a cold field-emission gun, where temperature is determined from thermal-expansion-induced shifts in the volume plasmon energy in accordance with the free-electron model. **Figure e** reproduced from ref. [36] under the terms of the CC-BY 4.0 license. Copyright 2019 The Authors. Published by Wiley-VCH GmbH. **Figure f** adapted from ref. [37] under terms of the CC-BY-NC-ND 4.0 license. Copyright 2025 The Authors. Published by American Chemical Society.

An overview of several widely used micro- and nanoscale thermometry approaches is presented in **Figure 1**. Each technique offers distinct advantages and limitations and differs in terms of spatial and temperature resolution (see **Table 1** for a detailed comparison). **Figure 1a** illustrates the working principle of IR thermography, an all-optical technique that retrieves surface temperature by detecting the thermal radiation naturally emitted by a sample. In accordance with Planck's law, the spectral radiance of this emission increases with temperature, enabling IR detectors to convert collected radiation into an electrical signal from which temperature is computed [38]. A typical IR thermography system comprises the emitting object, the propagation medium (air or vacuum), and an optical–detector module that focuses the emitted IR radiation onto a broadband or spectral-band detector. Although IR thermography is a highly mature and robust technology, its applicability at the micro- and nanoscale is fundamentally constrained. It is limited to a spatial resolution of ~10 µm, and accurate measurements require careful control of emissivity variations and background reflections. **Figure 1b** depicts how time-domain thermoreflectance (TDTR), a pump–probe ultrafast optical thermometry technique, is used to measure thermal properties such as thermal conductivity and heat capacity [39]. In this method, the sample is first coated with a thin metallic transducer layer and heated by a pump laser pulse, while a time-delayed probe pulse detects the resulting change in reflectivity (thermoreflectance) as the heated region cools [40, 41]. The experimental data, expressed as the ratio of in-phase to out-of-phase lock-in voltages, is plotted as a function of pump–probe delay time [39]. The decay of this ratio curve reflects how quickly heat diffuses into the sample and across interfaces, enabling the extraction of different thermal properties of the sample by comparing it with a thermal model. However, despite being a highly mature technique, TDTR remains poorly suited for biological samples due to its low biocompatibility. **Figure 1c** illustrates the fundamentals of Raman thermometry, in which temperature is determined from laser-induced Raman spectra by analyzing peak shifts, linewidth variations, or the anti-Stokes/Stokes intensity ratio [42, 43]. Raman scattering is an inelastic light–matter interaction in which incident photons excite a molecule from its ground state to a virtual energy state. If the molecule relaxes back to a different vibrational level, the scattered photon experiences a frequency shift, producing a Stokes Raman line. Conversely, if the molecule is already in an excited vibrational state, the interaction can generate an anti-Stokes line as the photon gains energy during scattering. Although the technique is fully non-contact, all-optical, and requires no sample preparation, it is comparatively slow and material-specific. **Figure 1d** illustrates another all-optical thermometry method, fluorescence thermometry, which exploits temperature-dependent changes in the fluorescence properties of certain materials to enable remote, high-sensitivity nanoscale thermal measurements. Fluorescent nanothermometers exhibit temperature-dependent responses arising from their molecular, atomic, and electronic-level dynamics [15]. These changes can be probed optically through shifts in fluorescence spectral position, variations in intensity or bandwidth, changes in lifetime, polarization anisotropy, or signals detected via optically detected magnetic resonance (ODMR). Finally, **Figures 1e** and **1f** present two non-optical thermometry methods commonly employed for nanoscale thermal sensing. **Figure 1e** presents a basic schematic of the SThM setup. In general, SThM employs a thermally active probe mounted on an AFM cantilever. Tip–sample interaction forces induce nanoscale deflections in the cantilever, which are detected either through a laser–mirror–photodetector system or via a piezoelectric cantilever, thereby enabling high-resolution thermal mapping [36, 44]. Some AFM-based thermocouple techniques measure the thermoelectric voltage generated at the tip–sample contact to map local temperature, eliminating the need for an AFM tip with an integrated thermal probe [45]. Although this method can achieve high spatial and temporal resolution, the complex heat transfer between the tip and the sample remains a major challenge, particularly because ballistic heat transport becomes significant at the nanoscale [46]. **Figure 1f** shows the temperature-dependent scanning of another non-contact sub-nanometer thermal imaging technique, plasmon energy expansion thermometry (PEET). PEET measures temperature-dependent shifts in the plasmon energy of a material [47]. As the sample thermally expands or contracts, the local electron density changes, thereby shifting the bulk plasmon energy detected by electron energy-loss spectroscopy (EELS). **Table 1** provides an overview of these thermometry techniques, summarizing their key advantages and limitations. The thermometry methods listed in this table are also compared using typical figures of merit, such as spatial and temperature resolution, which are introduced and discussed in the following section (**Section 3**).

**Table 1.** Summary of the advantages and limitations of high-resolution micro/nanoscale thermometry techniques. The table is adapted from the work published by Brites *et al.* [13].

| Systems | Typical Resolution | | Advantages | Limitations | References |
|---|---|---|---|---|---|
| | dx (μm) | dT (K) / dt (μs) | | | |
| Infrared Thermography | 10 | $10^{-1}$/10 | ▪ Non-contact and remote temperature measurement.<br>▪ Commercially mature and widely implemented technology.<br>▪ Provides full-field 2D/3D surface temperature mapping.<br>▪ Enables rapid, real-time thermal imaging. | ▪ Comparatively lower spatial resolution (~10 μm–mm)<br>▪ Requires accurate emissivity knowledge and calibration<br>▪ Sensitive to surface conditions and environmental reflections | [48-50] |
| Time-domain thermoreflectance | $10^{-1}$ | $10^{-2}/10^{-1}$ | ▪ Reliable non-contact method to read thermal conductivity and interface conductance for many bulk and thin-film samples.<br>▪ Comparatively simple sample preparation requirements | ▪ Diffraction-limited spatial resolution<br>▪ Needs smooth, reflective surfaces<br>▪ Requires an optically opaque transducer layer<br>▪ Transducers tend to fail at high temperatures.<br>▪ Accurate knowledge of the heat capacity of the material is required, which can be challenging for novel materials. | [39, 51-53] |
| Raman Thermometry | $10^{-1}$ | $10^{-1}/10^{6}$ | ▪ Label-free, non-contact and high spatial resolution method<br>▪ No sample preparation needed<br>▪ Anti-Stokes Raman scattering is widely used for cellular-level temperature measurements. | ▪ Weak Raman signal<br>▪ Extremely time-consuming process<br>▪ Crosstalk with fluorescence signal<br>▪ Material-specific technique; generally ineffective for most metals. | [54-58] |
| Fluorescence Thermometry | $10^{-1}$ | $10^{-2}/10$ | ▪ Nanoscale spatial resolution<br>▪ Compatible with the biological system | ▪ Calibration is challenging.<br>▪ Background fluorescence interferes with the signal detection<br>▪ Possible phototoxicity and photobleaching | [25, 59] |
| Scanning Thermal Microscopy (SThM) | $10^{-2}$ | $10^{-2}/10^{2}$ | ▪ Nanoscale spatial resolution<br>▪ Fully integrated with electronic devices | ▪ Contact-based slow acquisition time<br>▪ Heat transfer between the sample and the tip<br>▪ Near field radiative heat transfer<br>▪ Not well adapted to imaging biological cells.<br>▪ Hardly used in liquid environments | [60, 61] |
| Plasmon Energy Expansion Thermometry (PEET) | $10^{-3}$ | $10^{-1}/10$ | ▪ True nanoscale temperature mapping<br>▪ Used to image nanoscale temperatures of nanoparticles, membranes, and two-dimensional materials. | ▪ Requires a vacuum and an expensive setup<br>▪ Difficulty measuring temperature in materials with small thermal expansion coefficients<br>▪ Not applicable for the bio system<br>▪ Not robust for long-term or practical deployment | [37, 62, 63] |

## 3. Figures of Merit of a Thermometer

The performance of a thermometer is commonly benchmarked using key parameters, including spatial resolution, temperature sensitivity, temperature resolution, repeatability, reproducibility, accuracy, and precision. Among these, relative thermal sensitivity and temperature resolution are the most reported figures of merit in optical thermometry, as they directly determine the practical applicability of a given temperature sensor and enable a fair and meaningful comparison with alternative measurement systems [64].

### 3.1 Sensitivity

Sensitivity refers to the smallest detectable change in the thermometric observable induced by a variation in temperature. In the literature, two types of sensitivity are commonly used to characterize thermometers: absolute sensitivity ($S_a$) and relative sensitivity ($S_r$). The absolute sensitivity, $S_a$, quantifies the absolute change in a measured optical parameter, $O$(e.g., peak position, linewidth, or intensity), per unit change in temperature and is defined as:

$$S_a = \frac{\Delta O}{\Delta T} \quad (1)$$

A high absolute sensitivity is advantageous for detecting small temperature variations. However, its value is closely tied to the specific sensing material, optical observable, and measurement setup. These dependencies limit its usefulness for comparing different thermometric systems. To enable an objective evaluation of thermometer performance irrespective of sensor nature, operating principle, or material system, the relative thermal sensitivity $S_r$ is therefore adopted as a normalized and more universal figure of merit and is defined as:

$$S_r = \frac{\frac{\Delta O}{\Delta T}}{O} \quad (2)$$

$S_r$ can be directly compared across different systems, as it is expressed in standardized units of $K^{-1}$ or % $K^{-1}$. A thermometer should ideally exhibit high relative sensitivity, enabling a pronounced response even to small variations in temperature.

### 3.2 Temperature Resolution

Temperature resolution is the smallest temperature variation that can be reliably detected above the measurement system's noise floor. As a relative metric, the temperature resolution can be improved by increasing the signal integration time, thereby reducing statistical noise. The temperature resolution is commonly defined as

$$\eta_T = \sigma_T \sqrt{t_m} \quad (3)$$

where $t_m$ is the integration time and $\sigma_T$ is the uncertainty of the temperature. The temperature uncertainty $\sigma_T$ originates from intrinsic thermal fluctuations associated with the finite size of the thermal sensor and can be expressed as

$$\sigma_T = T\sqrt{\frac{k_B}{VC_v}} \quad (4)$$

where $k_B$ is the Boltzmann constant, $V$ is the volume of the quantum sensor, and $C_v$ is the volumetric heat capacity. A smaller temperature resolution corresponds to superior thermometric performance.

In some studies, temperature resolution is defined as the smallest temperature change that produces a detectable variation in the measured signal and is sometimes referred to as temperature uncertainty [64]. It is typically expressed in kelvin (K) and evaluated as the ratio of the measurement uncertainty (represented by the standard deviation) to the sensitivity.

$$\Delta T_{min} = \frac{\sigma}{S_a} \quad (5)$$

Here, $\sigma$ represents the standard deviation of the parameter used for temperature estimation.

In time-resolved measurements, the temporal resolution ($\Delta t_{min}$) is defined as

$$\Delta t_{min} = \frac{\Delta T_{min}}{|dT/dt|} \quad (6)$$

It corresponds to the minimum time interval over which the change in temperature does not exceed $\Delta T_{min}$. Temperature resolution indicates the smallest detectable temperature change, whereas temporal resolution represents the minimum timescale over which temperature variations can be resolved.

## 4. Fluorescence Thermometry

One of the most promising and precise approaches for remotely measuring temperature at the micro- and nanoscales is luminescence thermometry, also known as fluorescence thermometry, which has evolved into a broad and rapidly advancing research field over the past decade. Although luminescence thermometry was first demonstrated by Neubert [65] during the development of fluorescent lamps, the field attracted little sustained attention for several decades thereafter. The development of this field can be broadly categorized into three distinct eras: an initial phase spanning the 1930s to the 1980s, followed by a second period extending from the 1980s through the 2000s, and a third, ongoing era that has evolved from 2000 to the present [66]. Neubert's concept was first experimentally demonstrated by Urbach *et al.* [67] in 1949, which subsequently laid the groundwork for early thermographic phosphor applications. Following the pioneering contributions of Neubert and Urbach, research activity in this area remained largely dormant until the early 1980s, with fewer than ten publications per year, as illustrated in **Figure 2a.** The first era is considered to end with the 1979 introduction of Luxtron's first commercial phosphor-based optical thermometer, which utilized a $Eu^{3+}$ phosphor-coated optical fiber and a ratiometric fluorescence readout, achieving a temperature resolution of 0.1 °C and, to date, remains the only commercially available product based on luminescence thermometry [68]. During the second era (1980–2000), significant advances in materials science and instrumentation emerged, including the development of new phosphor materials and the introduction of laser-based techniques, leading to refined methodologies and expanded applications, particularly in industrial and scientific contexts. A major advance occurred in the early 2000s with the emergence of luminescent molecules and nanoparticles doped with TM and $Ln^{3+}$ ions, enabling microscale temperature sensing [69, 70]. These systems enabled non-contact, spatially resolved thermal measurements and marked a turning point in the development of luminescence thermometry. The application of fluorescence thermometry to diverse fields reignited interest in the field, driving rapid growth between 2006 and 2012, with approximately 100 publications per year (**Figure 2a**) and citations increasing from ~1000 to ~3000 (**Figure 2b**). Following 2012, the field expanded rapidly and diversified, with annual publications approaching ~1200 and yearly citations reaching ~45000 by the end of 2025. This growth marks the transition of luminescence nanothermometry into a mature, forward-looking research domain. While new classes of thermographic phosphors continue to emerge, current and future efforts increasingly emphasize establishing robust theoretical foundations and standardized measurement protocols, along with rigorous evaluation of reliability, repeatability, and reproducibility. Looking ahead, the integration of artificial intelligence for advanced data analysis and the exploration of application-driven frontiers are expected to play a central role in shaping the next phase of the field [71]. As a result, fluorescence thermometry has evolved into a highly versatile platform with applications spanning nanoelectronics, nanophotonics, micro- and nanofluidics, catalysis, and cell biology, as illustrated in **Figure 2c**. Fluorescence thermometry employs a wide range of phosphors, including color centers in diamond, QDs, UCNPs, and organic dyes. Among these, diamonds, QDs, and UCNPs are the most widely used across various applications. These probes are discussed in detail in the following subsections.

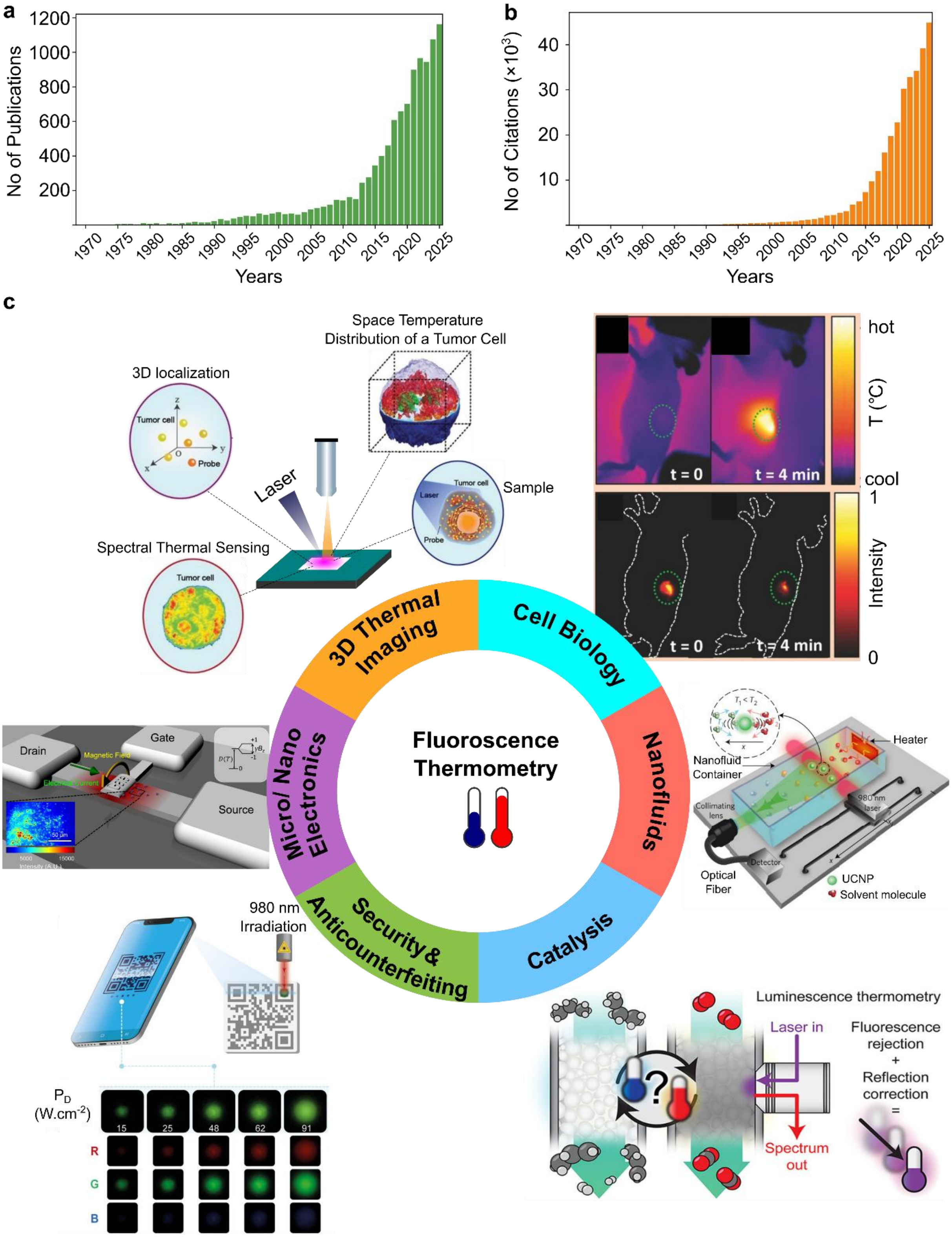

**Figure 2.** Bibliometric analysis and applications of fluorescence nanothermometry **(a)** Annual number of publications and **(b)** annual number of citations for scientific papers retrieved from the Web of Science (Clarivate Analytics), Core Collection (1900–2025). The search query was: *(((luminescence OR luminescent OR fluorescence OR fluorescent) AND (thermometry OR*

*nanothermometry OR thermometer* OR nanothermometer*)) OR “phosphor thermometry” OR “thermographic phosphors” OR ((diamond OR microdiamond* OR nanodiamond*) AND (thermometry OR nanothermometry OR thermometer* OR nanothermometer*))). The search was conducted on March 19, 2026. **(c)** Overview of the broad spectrum of applications enabled by fluorescence nanothermometry, spanning cell biology [Reproduced with permission [72]. Copyright 2016, John Wiley and Sons], three-dimensional thermal imaging [Adapted with permission [73]. Copyright 2021, John Wiley and Sons], nanoelectronics [Reproduced with permission [74]. Copyright 2020, American Chemical Society], catalysis [Reproduced from ref [75] under the terms of the CC-BY 4.0 license. Copyright 2025, The Authors. Published by American Chemical Society], security and counterfeiting [Reproduced from ref [76] under the terms of the CC-BY 4.0 license. Copyright 2021, The Authors. Published by Wiley-VCH GmbH], and nanofluidics [Reproduced with permission [77]. Copyright 2016, Springer Nature].

## 5. Diamond-Based Fluorescence Thermometry

Among various fluorescent phosphors, optically active color centers in the diamond lattice are highly promising thermal sensors. This is due not only to diamond’s nanoscale size, wide bandgap, and exceptional photophysical properties, but also to its ultra-high thermal conductivity, excellent mechanical and chemical stability, and inherent biocompatibility [78, 79]. Owing to the exceptional rigidity of its crystal lattice, diamond enables highly efficient phonon transport, making it one of the most thermally conductive materials found in nature. This outstanding thermal conductivity renders diamond particularly well suited for applications in heat management and high-precision temperature sensing. Nitrogen-vacancy (NV) centers, along with Group-IV color centers such as the germanium-vacancy (GeV), silicon-vacancy (SiV), tin-vacancy (SnV), and lead-vacancy (PbV), are among the most widely studied diamond color centers for temperature sensing. Broadly, diamond-based thermometry can be classified into two categories: spin-based diamond thermometry and all-optical diamond thermometry. The first method relies on ODMR, which measures shifts in the defect's spin resonance frequency. These shifts originate from the temperature dependence of the zero-field splitting of the ground state in defects with $C_{3v}$ symmetry [80]. The second method is an all-optical approach that exploits temperature-induced variations in the photoluminescence (PL) properties of the color center. Changes in parameters such as the zero-phonon line (ZPL) position, emission intensity, linewidth, or fluorescence lifetime provide temperature-sensitive optical signatures without the need for microwave excitation.

## 5.1 Spin-based diamond thermometry

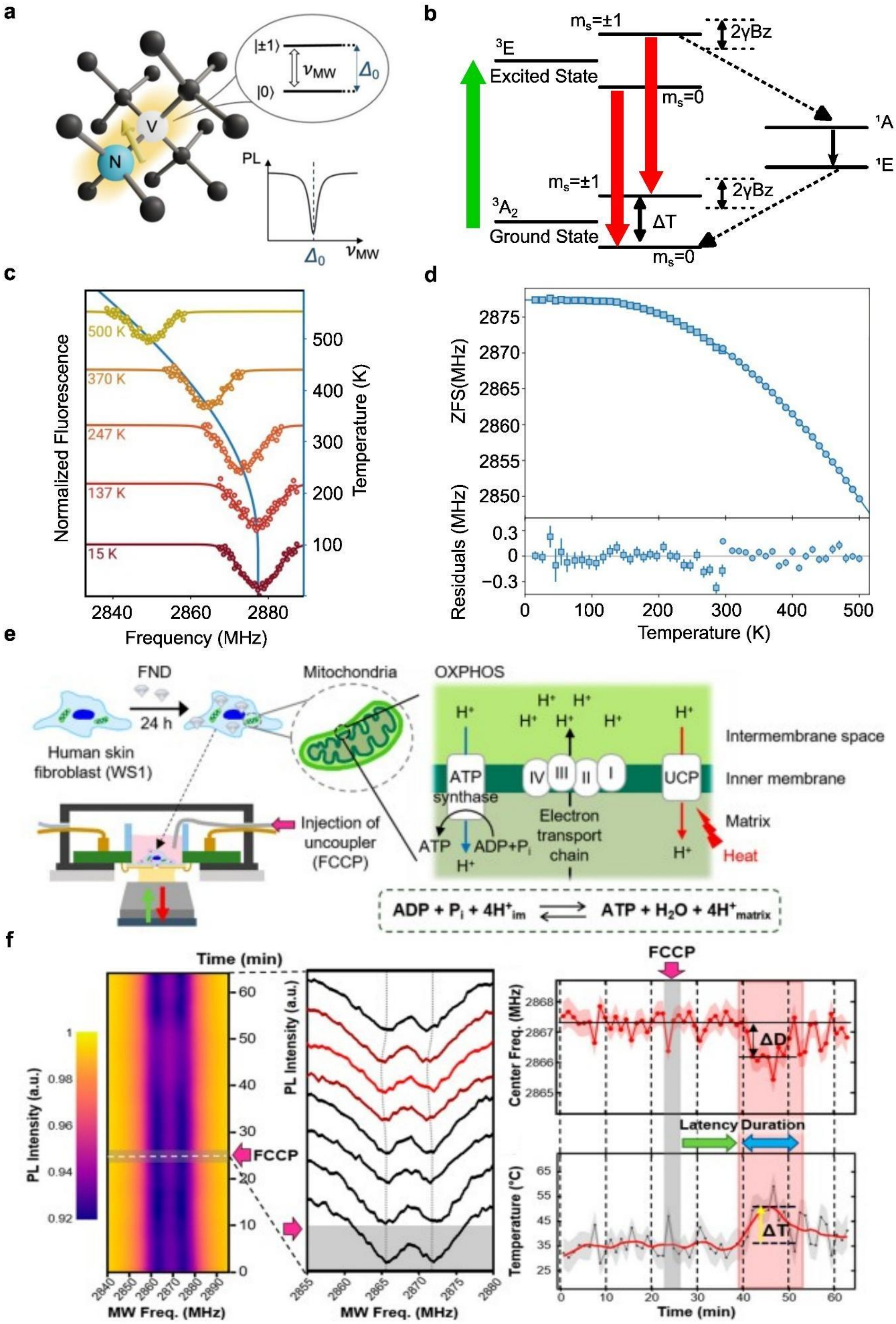

**Figure 3.** NV-based ODMR thermometry **(a)** Conceptual depiction of an NV center embedded in the diamond crystal, where carbon atoms are shown in black, the substitute nitrogen atom in blue, and the lattice vacancy in white; the associated electronic spin orientation is indicated by a yellow arrow. The diagram also presents the electronic structure of the triplet ground state, highlighting the zero-field splitting $\Delta_0$ . The lower-right panel displays a typical photoluminescence response obtained under microwave (MW) frequency modulation, demonstrating the principle of optically detected magnetic resonance. Reproduced from [81] under the terms of the CC-BY 4.0 license. Copyright 2021, AIP Publishing. **(b)** Simplified energy-level diagram of the NV center in diamond. **(c,d)** Temperature-dependent behavior of ODMR and Temperature dependence of the zero-field splitting parameter. Data points show the mean zero-field ODMR center frequency, averaged over multiple individual NV centers at each temperature, together with the corresponding fit residuals. Reproduced with permission [82]. Copyright 2023, American Physical Society. **(e)** Schematic depiction of intracellular temperature dynamics in fibroblast cells induced by ATP synthesis inhibition. Reproduced with permission [83]. Copyright 2025, American Chemical Society. **(f)** Electron spin resonance (ESR) spectra with Lorentzian fits, used to extract the center frequency, from which the temperature is determined using the ΔD/ΔT calibration. Reproduced with permission [83]. Copyright 2025, American Chemical Society.

The concept of diamond quantum thermometry was experimentally demonstrated in 2013, when three independent groups reported temperature sensing based on NV centers, marking a milestone toward nanoscale thermal mapping across various application domains [84-86]. NV centers are highly promising probes for nanoscale temperature sensing, owing to their exceptional sensitivity to temperature-induced lattice strain and their excellent biocompatibility [87]. The NV center is a point defect in diamond consisting of a substitute nitrogen atom adjacent to a lattice vacancy (a missing carbon atom), as shown in **Figure 3a**. It possesses $C_{3v}$ symmetry, with the threefold symmetry axis oriented along the N–V direction. In its negatively charged state, formed by capturing an additional electron from the lattice, the NV center exhibits spin-triplet ground and excited electronic states, as illustrated in **Figure 3b**. A simplified working model of the NV energy-level structure, including the triplet ground and excited states and the intermediate metastable singlet states, is shown in **Figure 3b**. The triplet ground-state ($^3A_2$) shows an axial zero-field splitting (ZFS) $D_{gs}$ of approximately 2.88 GHz at room temperature between the $m_s = 0$ and $m_s = \pm1$ sublevels, primarily originating from spin–spin interactions. The excited triplet state ($^3E$) gives rise to broadband photoluminescence with a zero-phonon line at 1.945 eV (637 nm) and possesses an orbital doublet character. When microwave radiation is tuned to the transition between the $m_s = 0$ and $m_s = \pm 1$ ground-state spin levels, a pronounced drop in fluorescence is observed—an effect known as ODMR, which forms the basis for sensitive spin-based sensing. Thermometry exploits the temperature dependence of the zero-field splitting parameter $D_{gs}$ to measure temperature changes. Three main measurement protocols are commonly used to extract temperature-dependent frequency shifts in the ODMR signal of NV centers: (i) continuous-wave (CW) ODMR spectral measurements, (ii) multipoint ODMR techniques, and (iii) pulsed ODMR methods [88]. The CW-ODMR method is the simplest approach for tracking zero-field splitting shifts, as it measures the full ODMR spectrum at different temperatures. Overall, CW-ODMR offers a simple and wideband approach, making it ideal for initial resonance identification and temperature sensing over broad ranges. A key limitation of CW-ODMR is its slow speed and reduced precision in NV ensembles, where strain-induced ODMR splitting increases fitting uncertainty and degrades temperature sensitivity [89]. The multipoint ODMR method builds on CW-ODMR but achieves much higher speed and practical precision by extracting the resonance shift from fluorescence measured at only a few selected microwave frequencies. The method significantly speeds up temperature measurements, enabling rapid data integration and high-precision, real-time thermometry. This technique has been demonstrated in several reported studies [84, 90-92]. On the other hand, pulsed ODMR methods achieve high sensitivity by suppressing spin decoherence effects. Wang *et al.* demonstrated that using a thermal Carr-Purcell-Meiboom-Gill (TCPMG-8) sequence extended the spin coherence time to 108 μs, resulting in an enhanced thermal sensitivity of 10.1 mK $Hz^{-1/2}$ [93]. In another study, Liu *et al.* demonstrated robust quantum coherence control of nitrogen-vacancy center electron spins in nanodiamonds at temperatures approaching 1000 K by combining room-temperature spin initialization and readout with pulsed-laser heating and rapid diffusion cooling on amorphous carbon films [94]. Recently, an analytical model was reported that captures the temperature dependence of the NV center zero-field splitting from 15 to 500 K, showing that the ZFS shift scales with the occupation numbers of two representative phonon modes [82]. **Figure 3c** shows pulsed ODMR spectra recorded at different temperatures, and **Figure 3d** presents the average ZFS measured for NV as a function of temperature, together with a fitted model of the form $D(T) = D_0 + \sum_{i=1}^{M} c_i\, n_i$. D(T) represents the ZFS at temperature T, while $D_0$ is the ZFS at 0 K. The model considers M discrete phonon modes, each indexed by $i$. Each coefficient $c_i$ quantifies how a given phonon mode affects the ZFS by altering atomic positions and vibrational amplitudes within the lattice. The average occupation number of the $i$-th phonon mode is denoted by $n_i$. Although demonstrated on single NV centers in bulk diamond, the proposed ZFS–temperature model is broadly applicable to NV ensembles, near-surface NVs, and nanodiamonds, owing to the sub-nanometer wavelengths of the relevant phonon modes. Spin-based thermometry offers powerful capabilities for precise intracellular temperature sensing. Lee *et al.* demonstrated real-time monitoring of temperature changes during adenosine triphosphate (ATP) synthesis and inhibition using NV-based spin thermometry as seen in **Figure 3e** [83]. The strong temperature dependence ($\Delta D/\Delta T \approx -77$ kHz °C$^{-1}$) enables precise temperature detection by monitoring shifts in the electron spin resonance (ESR) frequency under continuous-wave microwave excitation and laser illumination, as illustrated in **Figure 3f**. Different spin-based thermometry studies are summarized in **Table 2**, with key figures of merit for the thermal sensor and their respective applications.

**Table 2. Comparison of ODMR-based diamond thermometry using different defects, summarizing ODMR modality, spatial resolution, temperature resolution, and target platforms/applications.**

| Defect | ODMR Type | Spatial Resolution | Temperature Resolution | Platform/Applications | References |
|---|---|---|---|---|---|
| NV | CW ODMR | Bulk Diamond (3×3×3 mm$^3$) | 10 mK√Hz | Diamond sensors attached to external objects / Handheld, non-contact ODMR thermometry platform | [95] |
| NV | CW ODMR | 100 nm | 11 mK√Hz | Hybrid FND–magnetic nanoparticle/ Point nanothermometry | [91] |
| NV | CW ODMR | 200 µm | 18 mK√Hz | Fiber-based quantum thermometer / Thermal imaging of electronic circuit | [96] |
| NV | CW ODMR | 20 µm | 22.9 mK√Hz | Fiber Optic Probe/Electronic Devices | [97] |
| NV | CW ODMR | Sub micrometer | 35 mK √Hz | Non-invasive extracellular single-cell thermometry platform | [98] |
| NV | CW ODMR | 250 µm | 63 mK√Hz | Fiber-based quantum thermometer/spatial temperature mapping of heated wire | [99] |
| NV | CW ODMR | 200 nm | 96±9 mK√Hz | Hydrogel-mediated hybrid ND–Ni MNP quantum sensing platform / Bio-chemical reaction | [100] |
| NV | CW ODMR | 30 µm | ~130 mK√Hz | Fiber-based thermometer microprobe platform/ Temperature sensing of the polyimide film surface | [101] |
| NV | CW ODMR | 100 nm | 1.4 K√Hz | In-vivo quantum thermometry platform for real-time temperature monitoring inside living organisms (*C. elegans*) | [92] |
| NV | CW ODMR | 100 nm | 2 K√Hz | ND on glass coverslip/ Measures Photothermal heating from nearby gold nanorods | [90] |
| NV | CW ODMR | 3.7 nm | 2.1 K √Hz | Intracellular thermometry | [102] |
| NV | CW ODMR | 1 µm | 2.7 K√Hz | NDs embedded in PDMS film /Thermal mapping of electronic circuit | [103] |
| NV | CW ODMR | 1 µm | 2.6°C per pixel | Dropcasted NDs/Temperature rise in Commercial GaN HEMT devices | [74] |
| NV | CW ODMR | ~50–60 nm | Sub-degree precision | In situ Cellular Thermometry | [104] |
| NV/SiV | NV ODMR and SiV spectroscopic shift) | Bulk Diamond | 22 and 86 mK√Hz | Hybrid cross-correlated optical thermometry /Thermal measurement at the nanoscale | [105] |
| NV/SiV | NV ODMR and SiV spectroscopic shift) | 11.2 ± 3.0 nm | 360 mK √Hz | Multimodal nanothermometry platform enabling simultaneous NV-based ODMR and SiV-based all-optical thermometry/ | [106] |
| | | | | Thermal measurement at nanoscale | |
| NV | Pulsed ODMR | 1 mm | 76 µK √Hz | Hybrid nanothermometer-diamond nanopillar coupled with a single magnetic copper-nickel alloy nanoparticle/ monitoring thermal dynamics in micro-scopic systems | [107] |
| NV | Pulsed ODMR | 1 µm/5 nm | 5 mK√Hz/ 130 mK√Hz | ND sensor/ Temperature measurement around thin metallic wires | [85] |
| NV | Pulsed ODMR | 200 nm | 9 mK√Hz | ND and Gold NPs into a human tissue/ Cell Biology | [84] |
| NV | Pulsed ODMR | 1 mm | 10.1 mK√Hz | Implanted NV center array in the surface layer of a diamond chip/ temperature distribution on the diamond chip surface | [93] |
| NV | Pulsed ODMR | ~140 nm | 250 mK√Hz | Nanodiamonds deposited on amorphous carbon films / High temperature thermometry | [94] |

### 5.2 All-optical diamond thermometry

The reliance on microwave excitation in spin-based diamond thermometry can limit its applicability in environments where microwave delivery or associated heating is challenging. This has motivated the development of all-optical approaches that measure temperature by monitoring changes in photoluminescence intensity, as well as shifts and linewidth variations of the ZPL. The simplicity of all-optical thermometry enables temperature sensing with various diamond color centers, particularly group IV centers (SiV, GeV, SnV, PbV), which possess $D_{3d}$ symmetry (See **Figure 4a**) and feature E-symmetric ground ($^2E_g$) and excited ($^2E_u$) states (**See Figure 4c**). Each electronic level is split by spin–orbit coupling. However, these spin-dependent sublevels are not suitable for temperature sensing at room temperature.

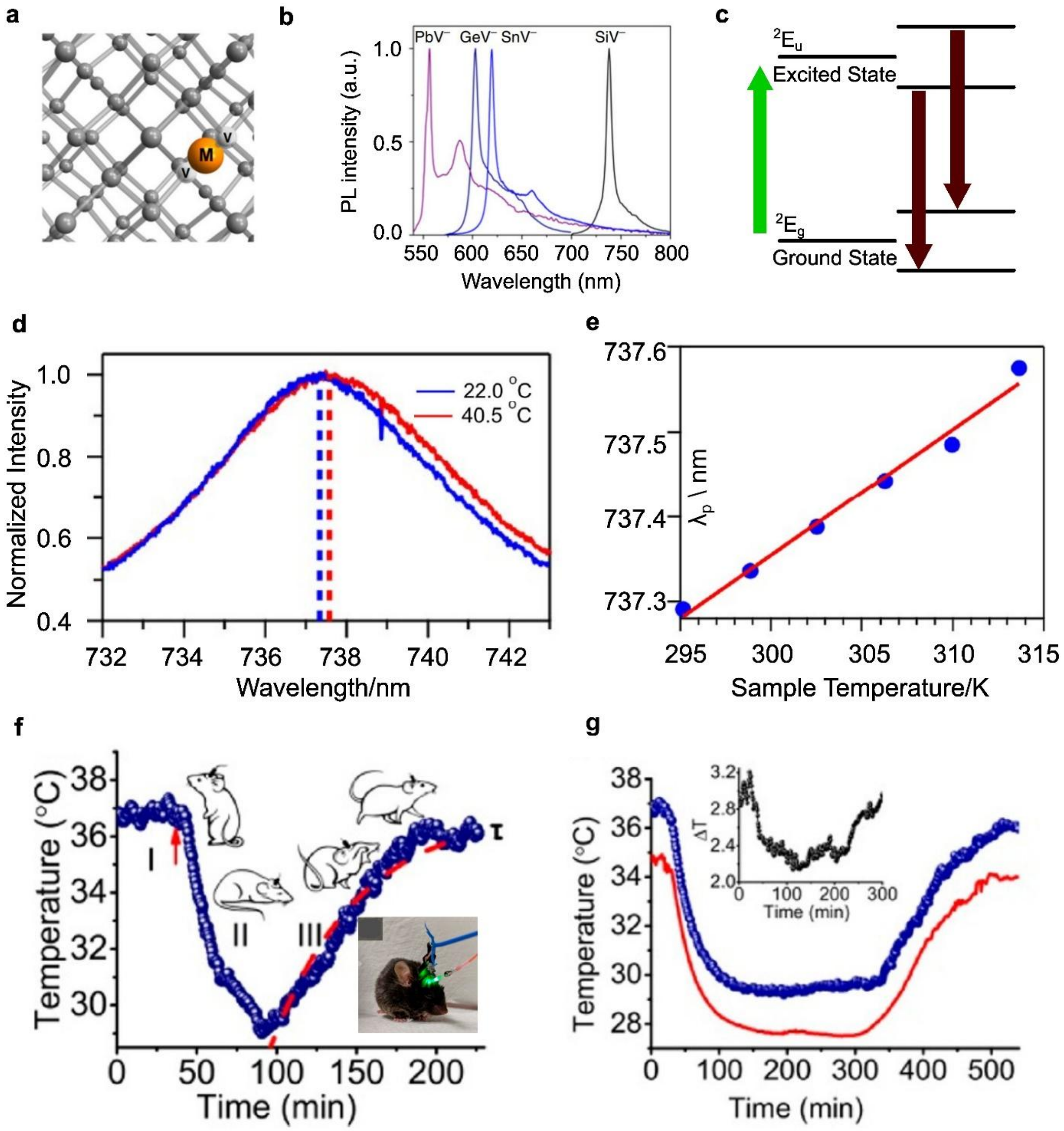

**Figure 4.** All optical diamond thermometry **(a)** Atomic structure of group IV color centers exhibiting $D_{3d}$ symmetry in a split-vacancy configuration, where the group IV impurity atom (M, orange) is positioned between two adjacent carbon vacancies (V, white). [Reproduced from [108] under the terms of the CC-BY 4.0 license. Copyright 2019, Springer Nature], **(b)** Representative photoluminescence (PL) spectra of $SiV^-$, $GeV^-$, $SnV^-$, and $PbV^-$ color centers at room temperature. [Reproduced from ref [108] under the terms of the CC-BY 4.0 license. Copyright 2019, Springer Nature], **(c)** Energy-level diagram of group IV color centers illustrating the principle of all-optical thermometry. **(d)** Photoluminescence spectra of SiV-containing nanodiamonds acquired at low and elevated temperatures, demonstrating a temperature-induced redshift of the ZPL. [Adapted with permission [109]. Copyright 2022,

Elsevier] (e) Dependence of the ZPL peak position ($\lambda_p$) on temperature. [Adapted with permission [109]. Copyright 2022, Elsevier] (f) In vivo brain thermometry using a fiber-coupled GeV diamond probe. Temporal evolution of brain temperature in a freely moving mouse measured with a GeV-diamond sensor attached to a fiber probe, showing normal conditions (I), drug-induced hypothermia (II), and recovery (III). [Adapted with permission [110]. Copyright 2020, American Chemical Society] (g) In vivo temperature comparison. Fiber-probe measurements from the motor thalamus (blue) are compared with subcutaneous type-K thermocouple readings (red) in a freely moving mouse. [Reproduced with permission [110]. Copyright 2020, American Chemical Society]

Consequently, thermometry based on group-IV color centers has been implemented exclusively using all-optical methods. Plakhotnik and co-workers were among the first to propose exploiting the temperature-dependent optical properties of NV centers—such as fluorescence intensity and excited-state lifetime over the 300–670 K range for thermometric applications [111]. An all-optical thermometry scheme based on an ensemble of SiV centers in diamond has been reported, utilizing the temperature-dependent shift of the zero-phonon line. A temperature resolution of 360 mK $Hz^{-1/2}$ was achieved in bulk diamond, while measurements using 200 nm nanodiamonds yielded 521 mK Hz−1/2, demonstrating sub-kelvin resolution in both cases. Choi *et al.* demonstrated an all-optical thermometry method using SiV centres, in which multiparametric analysis of the full fluorescence spectral profile was employed to determine temperature. This approach achieved an impressive temperature resolution of ~13 mK $Hz^{-1/2}$—representing nearly a three-orders-of-magnitude improvement in readout speed—and enabled temperature measurements over sub-250 nm spatial dimensions. Zhang *et al*. introduced an ultrasensitive all-optical nanothermometry platform based on SiV nanodiamonds, employing a two-photon excitation scheme to efficiently excite SiV centers. This nonlinear excitation approach enabled high-precision temperature sensing with an exceptional resolution of 6.6 mK $Hz^{-1/2}$, which, at the time of publication, represented the highest reported resolution for all-optical SiV-based nanothermometry. In another study, air-oxidized and polyglycerol-functionalized SiV-containing detonation nanodiamonds with an average diameter of ~20 nm—representing the smallest nanodiamonds reported for color-center-based thermometry—were investigated [109]. The authors observed a linear redshift of the SiV ZPL with increasing temperature in the 22.0–40.5 °C range, demonstrating their effectiveness for nanoscale optical thermometry, as shown in **Figure 4d–e**. An alternative all-optical diamond quantum thermometry strategy has been reported, utilizing anti-Stokes excitation of GeV color centers to enable temperature sensing [112]. In a recent study, SnV centers were also characterized over a wide temperature range from cryogenic to room temperature, achieving a precision of 500 mK $Hz^{-1/2}$ [113]. The linear temperature dependence of both the ZPL shift and the linewidth further supports robust, self-consistent temperature readout. All-optical thermometry has emerged as a promising approach for biological applications due to its microwave-free operation. In vivo brain temperature monitoring was recently realized using a fiber-coupled GeV diamond probe, demonstrating an accuracy better than 0.15 °C and temperature resolution of 230 mK $Hz^{-1/2}$ [110]. **Figure 4f and Figure 4g** show fiber-probe-based brain thermometry in a freely behaving mouse, where the temperature is extracted from the ZPL-to-phonon side band (PSB) energy ratio, quantified through the Debye–Waller factor (DWF). In vivo brain temperature was monitored in awake, freely moving mice in their home cage, concurrently with control measurements from a subcutaneously implanted type-K thermocouple on the head. In biological optical sensing, the NIR-II spectral range (1000–1700 nm) is significantly more advantageous than the shorter emission wavelengths of most diamond color centers, as photons in this window experience reduced tissue scattering, lower absorption by biological chromophores, and diminished autofluorescence, thereby enabling deeper penetration and higher signal-to-noise ratios. A recently identified defect in Si-doped diamond emits at 1221 nm in the O-band, within a key biological transparency window [114]. This center exhibits pronounced temperature dependence above 150 K, enabling accurate sensing up to ~420 K. Temperature can be extracted from ZPL spectral shifts, linewidth broadening, or the ZPL-to-PSB intensity ratio, achieving high-resolution thermometry. Temperature-dependent near-infrared fluorescence of nickel-related color centers in diamond has likewise been studied [115]. Notably, the ZPL wavelength falls within the biological transparency window, highlighting its potential for bio-optical sensing applications. An overview of representative all-optical diamond thermometry studies is provided in **Table 3**, where key performance parameters and application areas are systematically compared.

**Table 3. Comparison of all optical diamond thermometry using different defects, summarizing optical observables, spatial resolution, temperature resolution, and target platforms/applications.**

| Defect | Optical Observable | Spatial Resolution | Temperature Resolution | Platform/Applications | References |
|---|---|---|---|---|---|
| NV | ZPL peak position | 30 µm | 16 mK√Hz | Fiber Optic Platform/local temperature monitoring in a freely moving animal | [116] |
| NV | ZPL Amplitude | 0.56 × 0.57 × 2.18 $\mu m^3$ | 80 mK√Hz | 3D thermal mapping/Biochemical reaction thermal readout | [117] |
| NV | DWF of ZPL | 30 nm | 100 mK√Hz | Suitable for in vivo biology applications where microwave excitation is prohibited | [118] |
| NV | ZPL peak position | 400 µm | 131.5 mK√Hz | Fiber Optic Platform/ Monitor temperature changes in microscale chemical reactions | [119] |
| NV | ZPL intensity ZPL background ratio | Less than 50 nm | 0.3 K√Hz | Nanodiamond sensor | [120] |
| NV | Fluorescence Lifetime | 100 nm | ~16–57 K√Hz | Single Nanodiamond temperature sensing/ AFM nanomanipulation for precise temperature monitoring | [121] |
| GeV | ZPL shift ZPL Linewidth | 25 µm | 20 mK√Hz | Fiber Optic Probe with diamond sensor / 2D thermal Imaging | [122] |
| GeV | ZPL peak position ZPL linewidth | Bulk microdiamond | 37.5 mK√Hz | Suitable for biological and in vivo thermometry | [123] |
| GeV | ZPL peak position | 20 µm and 4 µm | 230 mK√Hz | Fiber Optic Platform/ Temperature measurements in the brain of freely behaving mice | [110] |
| GeV | ZPL peak position ZPL Linewidth | Bulk Microdiamond | 300 mK√Hz | Fiber Optic Probe with diamond sensor/ compatible with in vivo biological applications | [124] |
| GeV | ZPL linewidth | 300-500 nm | 4 K√Hz | Cryogenic Nanothermometry/ Microelectronic Devices | [125] |
| SiV | ZPL Intensity | 50 nm | 6.6 mK√Hz | Nanodiamond sensor/ Intracellular temperature measurements | [126] |
| SiV | Multiparametric | 70 nm | 13 mK√Hz | Drop cast on cover glass/ suitable for intracellular thermometry | [127] |
| SiV | ZPL peak position | 200 nm | 521 mK√Hz / 360 mK√Hz | Individual nanodiamonds as local probes/ suitable for nanoscale temperature sensing | [128] |
| SiV | Ratiometric (PL Intensity/Backscattered Laser) | 7.5 µm | 0.72 K√Hz | Fiber-based thermal sensor/ Real-time temperature monitoring in microelectronics | [129] |
| SiV | ZPL peak position and bandwidth | Micrometer size | 1.29 K√Hz | Silicon Pillar topped with microdiamond sensor/Thermal conductivity change of hydrogels during the gelation process is monitored | [130] |
| SiV | ZPL peak position | 20 nm | 2.9±1.3 K√Hz | Polyglycerol-coated Detonation nanodiamond sensor/ Suitable for Biological Applications | [109] |
| $SiV_2:H^{(-)}$ | ZPL peak position ZPL linewidth PSB/ZPL intensity ratio | ~200 nm | ≈ 0.57 K√Hz (from ZPL linewidth) ≈ 0.63 K√Hz | NIR thermometry platform/ Non-invasive temperature measurement in Microelectronics And biological tissue | [114] |
| | | | (from ZPL peak shift | | |
| SnV | ZPL peak position | - | 500 mK√Hz | Suitable for biological applications | [113] |
| GeV/SiV | Antistokes/ Stokes Ratio | 300−500 nm | 0.39 K√Hz | An array of nanodiamond sensors/ Real-time Temperature monitoring in Microelectronics | [131] |
| NV/SiV | Multiparametric | 1 µm | - | Bioimaging platform for in vitro temperature monitoring | [132] |
| Ni-related defect | ZPL Amplitude ZPL linewidth | Bulk microdiamond | 338 mK√Hz | NIR thermometry platform/ Suitable for Biological Applications | [115] |
| Ni-related defect | Luminescence peak intensity and Lifetime | 30 µm and 4 µm | - | NIR thermometry platform/ Suitable for Biological Applications | [133] |

Diamond-based thermometry can be affected by self-absorption and scattering from the surrounding environment, leading to changes in photoluminescence intensity and spectral position. However, some approaches rely on single-photon emission from individual color centers, which are largely immune to such effects [112]. Another important consideration is local heating of the diamond nanoprobe, arising from laser excitation or phonon-mediated nonradiative processes. In nanodiamonds and microdiamonds, this effect is often exacerbated by impurities, surface carbon, and the underlying substrate's thermal conductivity, leading to increased heating [134, 135]. In addition, thermometry based on NV centers often requires microwave excitation, which can be strongly absorbed in biological environments, leading to measurement artifacts. Finally, due to their nanoscale size, nanodiamonds are also sensitive to surface effects, such as interactions with ligands, surfactants, and trap states. Therefore, careful control experiments (e.g., in situ and ex situ) are necessary to distinguish temperature-induced changes in photoluminescence from environmental influences.

## 6. QDs-Based Fluorescence Thermometry

While diamond-based color centers offer exceptional photostability and high sensitivity, their implementation often requires relatively complex instrumentation and material processing. Consequently, semiconductor quantum dots have been explored as an alternative platform for optical thermometry. QDs, semiconductor nanocrystals typically 1–10 nm in size, exhibit size-tunable photoluminescence arising from quantum confinement, along with high brightness and facile integration, making them versatile probes for temperature sensing. QDs exhibit high brightness (PLQY ≈ 80–100% and luminance often exceeding $10^4$ cd/m² in QLEDs), excellent spectral tunability across the visible and sometimes into the NIR while maintaining narrow emission bandwidths (FWHM ≈ 15–40 nm), strong resistance to photobleaching, and, after appropriate surface functionalization, favorable biocompatibility for biological applications[136, 137]. QDs are typically based on II–VI, III–V, and IV–VI semiconductors and have been extensively synthesized from materials such as ZnS, CdS, ZnSe, CdTe, InP, and PbSe, enabling tunable emission across the UV-infrared spectrum, as illustrated in **Figure 5a**. In nanothermometry, Cd-core/ZnS-shell QDs are widely used to improve optical stability and biocompatibility [13]. The ZnS shell reduces nonradiative surface recombination, enhancing quantum yield and signal-to-noise ratio, while simultaneously isolating the potentially toxic core for safer bio applications. For thermometry applications, temperature influences the emission wavelength, intensity, and photoluminescence lifetime of quantum dots, allowing any of these parameters to function as a reliable temperature indicator. For instance, a temperature-dependent PL lifetime behavior—exhibiting a positive correlation with temperature—was reported in DDAB/TOAB-treated $CsPbBr_3$ QD films as shown by **Figure 5b** [138]. To further improve their thermometric performance, $CsPbBr_3$ QDs were subsequently embedded in polymer matrices, including polyethylene oxide (PEO), polymethyl methacrylate (PMMA), and polystyrene (PS). Conversely, a separate study explored the temperature-dependent properties of CdTe quantum dots (2.3–3.1 nm) incorporated into a NaCl protective matrix over 80–360 K [139]. A negative temperature dependence is observed, as PL lifetime analysis reveals that the intrinsic radiative lifetime of CdTe QDs exhibits minimal sensitivity to temperature, whereas thermally activated nonradiative recombination predominantly accounts for the quenching of PL intensity.

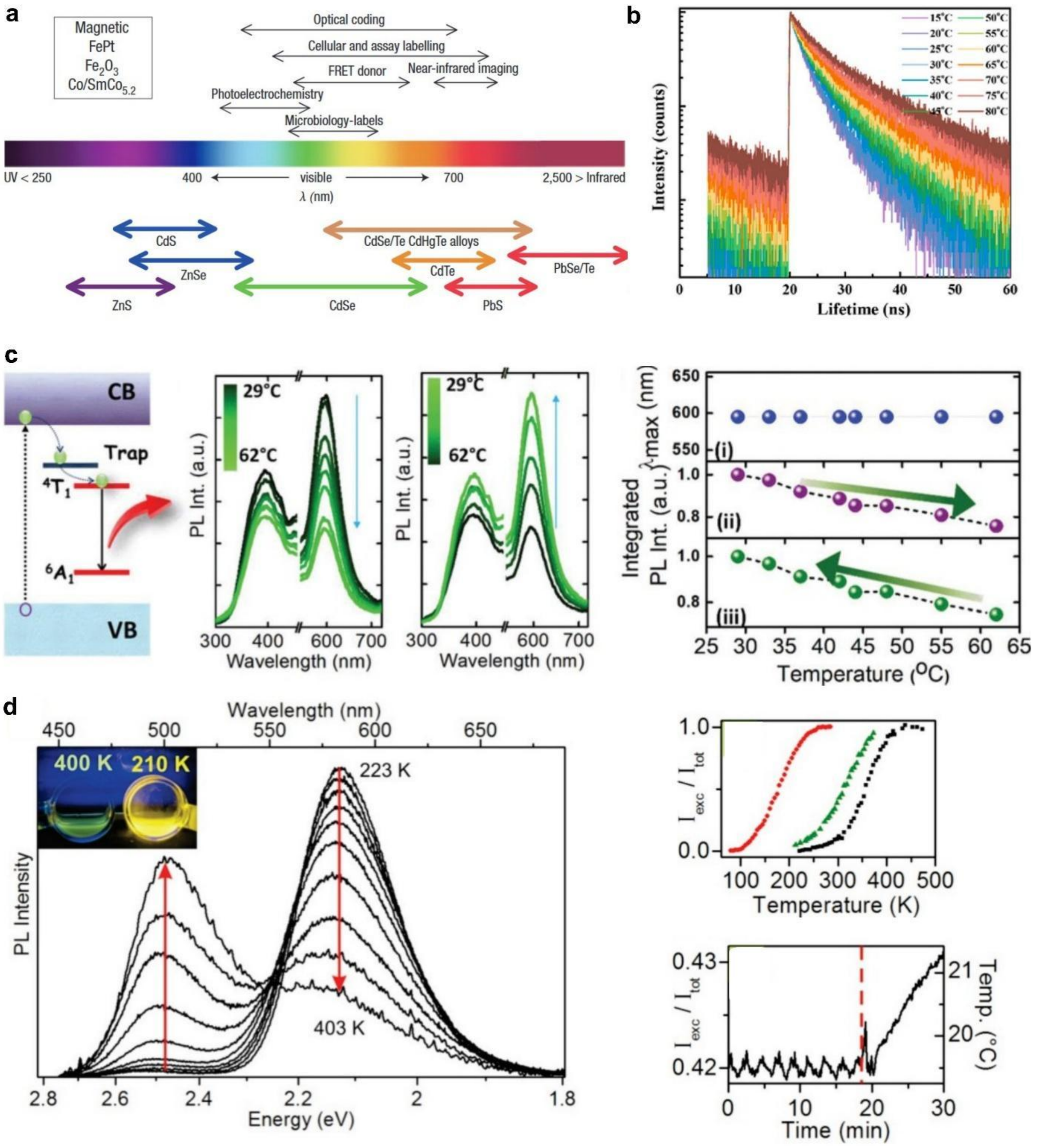

**Figure 5.** QDs-based fluorescence thermometry **(a)** Representative QD core materials plotted according to their emission wavelengths, highlighting their potential biological application windows. [Reproduced with permission [140]. Copyright 2005, Springer Nature] **(b)** Time resolved photoluminescence decay curves of didodecyl dimethylammonium bromide (DDAB)/tetraoctylammonium bromide (TOAB) capped $CsPbBr_3$ QD film at different temperatures. [Reproduced with permission [138]. Copyright 2024, Elsevier] **(c)** Schematic energy band diagram of $Mn^{2+}$-doped ZnS quantum dots (QDs), along with temperature-dependent photoluminescence (PL) spectra recorded during heating and cooling cycles. The variation of the PL peak position and the integrated emission intensity as a function of temperature is also presented. [Reproduced with permission [141]. Copyright 2021, Royal Society of Chemistry (RSC)] **(d)** Dual-emissive nanocrystals for optical thermometry. Temperature-dependent photoluminescence spectra of colloidal $Zn_{1-x}Mn_xSe$–ZnCdSe nanocrystals, acquired at 20 K intervals and normalized to the total integrated intensity, reveal a temperature-driven redistribution of emission intensity between the $Mn^{2+}$ and excitonic bands. [Reproduced with permission [142]. Copyright 2010, American Chemical Society]

In another study, a carbon QD-based thermometer was developed for intracellular thermometry, employing intensity-, ratiometric-, and lifetime-based detection schemes to validate temperature measurements at the cellular level [143]. Both the emission intensity and fluorescence lifetime decreased with increasing temperature. A major challenge moving forward is to enhance spatial resolution and realize accurate, quantitative temperature measurements at the single-pixel scale. A smartphone-assisted optical thermometry platform has also been reported in ref. [141], utilizing the PL quenching behavior of chemically synthesized $Mn^{2+}$-doped ZnS quantum dots (MZQDs, ~5 nm) and operating over a high-temperature range from room temperature to 145 °C. Two distinct emission bands were observed in the MZQDs at approximately 400 nm and 595 nm. The higher-energy emission (~400 nm) is attributed to zinc vacancies or surface-related defect states, while the lower-energy band (~595 nm) originates from the $Mn^{2+}$ internal 3d transition from the excited $^4T_1$ state to the ground $^6A_1$ state, as illustrated in **Figure 5c (left panel)**. Temperature-dependent PL spectra were recorded at various temperatures, as shown **in Figure 5c**. As the temperature increased from 29 °C to 62 °C, the PL intensity decreased markedly, while the emission peak position remained unchanged. Ratiometric optical thermometry at elevated temperatures has been reported in a separate study [142], employing $Mn^{2+}$-doped semiconductor nanocrystals based on $Zn_{1-x}Mn_xSe$/ZnCdSe core–shell structures, as shown in. These nanocrystals display strong temperature-dependent luminescence originating from two distinct emission bands, enabling self-referenced ratiometric thermometry that minimizes signal instabilities by simultaneously affecting both the measured and reference emissions. As illustrated in **Figure 5d**, temperature is determined from the ratio of the excitonic emission (higher energy) to the $Mn^{2+}$ emission (lower energy), which varies systematically with temperature. Recently, a $CuInS_2$/ZnS core–shell quantum dot nanothermometer was demonstrated using four temperature-dependent luminescence parameters—PL intensity, peak shift, lifetime, and excitation-based ratiometric intensity [144]. By combining these metrics via multiple linear regression, an outstanding sensitivity of up to 23% °C$^{-1}$ was achieved, among the highest reported for QD-based thermometers.

QD-based thermometry presents several limitations that must be carefully considered. QDs are prone to photobleaching and photo blinking, which can compromise signal stability and long-term measurements. Like nanodiamonds, they can also experience self-absorption and scattering from the surrounding environment, which can affect spectral accuracy. Due to their small size and high surface-to-volume ratio, surface-related effects are significant; thus, observed fluorescence quenching must be carefully attributed to temperature changes rather than chemical reactions at the surface. Additionally, thermal quenching in many QD systems is not fully reversible, raising concerns regarding measurement repeatability and long-term reliability[145].

## 7. UCNPs-Based Fluorescence Thermometry

While QDs provide strong and tunable fluorescence signals, their susceptibility to photobleaching, limited penetration depth, and background autofluorescence in biological environments have motivated continued interest in upconversion nanoparticles, which offer superior photostability, near-infrared excitation for deeper tissue penetration, suitability with cellular thermal sensing, and significantly reduced background interference [146-148]. UCNPs are a class of nanoscale fluorescent materials that convert low-energy photons into higher-energy emissions. Specifically, they absorb near-infrared light and emit at shorter wavelengths, such as in the visible region, through a photon upconversion process. UCNPs typically consist of a crystalline host matrix—commonly fluorides, oxides, phosphates, or sulfides of metal ions—doped with at least two different types of lanthanide ions [149]. In a typical configuration, one lanthanide ion serves as a sensitizer, absorbing excitation light and transferring the energy to a second dopant, the activator, which subsequently emits luminescence [150]. Depending on the dopant configuration, upconversion emission can be classified as single-center or multicenter, arising from either a single type of $Ln^{3+}$ ion or cooperative interactions between multiple $Ln^{3+}$ ions [12]. The most widely studied upconversion systems employ $NaYF_4$ as the host crystal, $Yb^{3+}$ as a sensitizer, and $Tm^{3+}$ as an activator. $Yb^{3+}$ is particularly effective due to its large absorption cross-section at 980 nm—a readily accessible wavelength—and its excited-state energy level, which closely matches those of $Er^{3+}$, $Ho^{3+}$, and $Tm^{3+}$, enabling efficient resonant energy transfer. Among these activators, $Er^{3+}$ is especially popular for its intense green emission and strong temperature dependence. In most UCNP-based thermometry approaches, temperature readout is achieved via a fluorescence intensity-ratio method based on Boltzmann thermometry, which requires thermally coupled energy levels with a small energy separation (typically <2000 $cm^{-1}$) to maintain thermal equilibrium. This condition is well satisfied in $Er^{3+}$, where the $^2H_{11/2}$ levels $^4S_{3/2}$ form a representative thermally coupled pair, enabling reliable ratiometric temperature sensing. Based on this, Sedlmeier et al. demonstrated that hexagonal $NaYF_4$ nanocrystals co-doped with $Yb^{3+}$ (sensitizer) and $Er^{3+}$/$Ho^{3+}$/$Tm^{3+}$ (activators) exhibit multiple emission bands with distinct temperature dependences, enabling ratiometric, self-referenced temperature readout in the physiologically relevant range of 20–45 °C with a resolution better than 0.5 °C [151]. In another study, one of the highest relative sensitivities ($S_r$) reported for upconversion-based thermometry (9.52% $K^{-1}$) was achieved using upconversion nanoparticles [152]. Under 980 nm excitation, the visible $Er^{3+}$ emission displayed pronounced temperature dependence across the 83–323 K range.

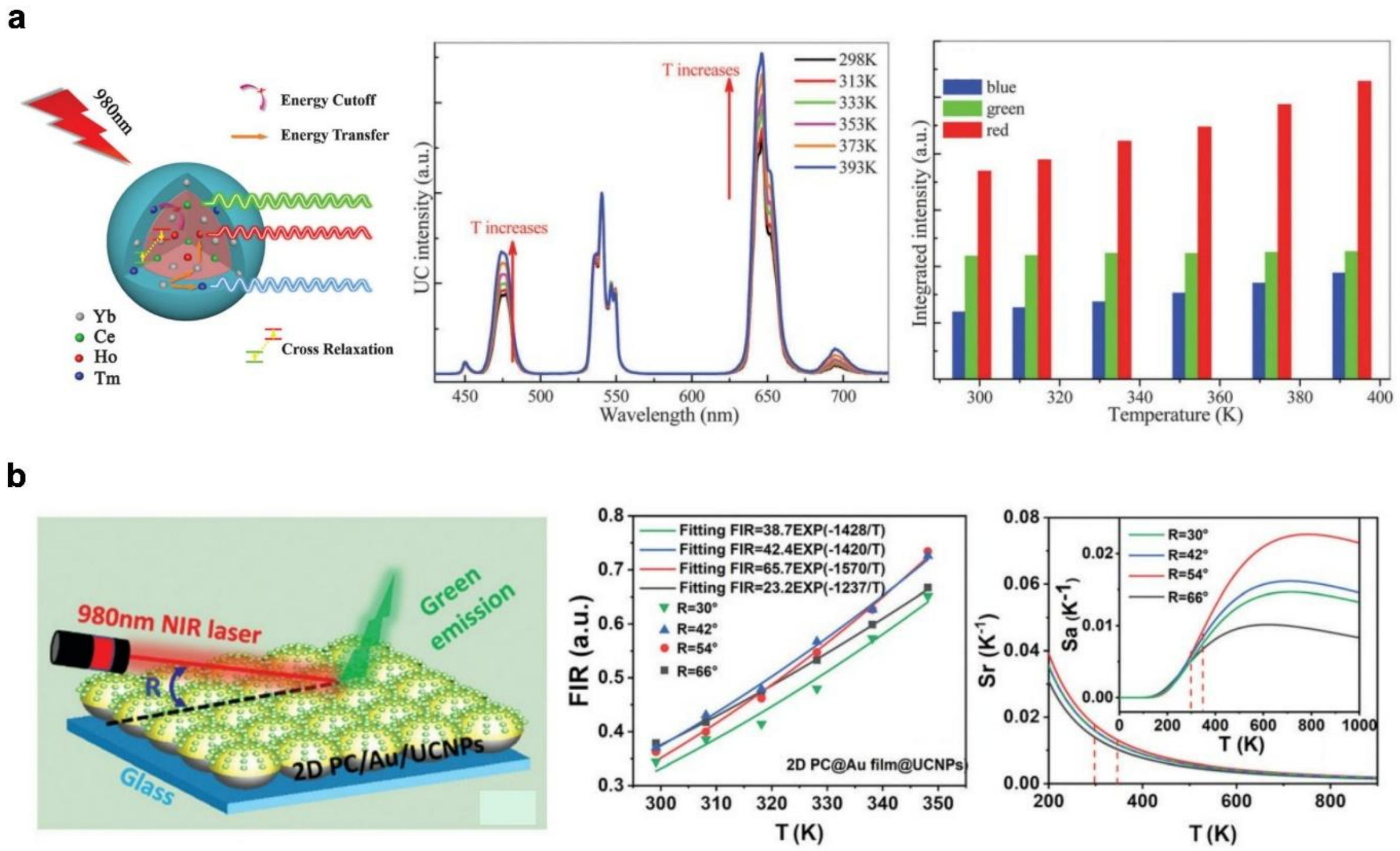

**Figure 6.** Upconversion nanoparticles (UCNPs) based fluorescence thermometry **(a)** Schematic illustration of the Yb/Ho/Ce:$NaGdF_4$@Yb/Tm:$NaYF_4$ core–shell nanostructure, along with temperature-dependent upconversion emission spectra recorded under 980 nm excitation and normalized to the green emission peak, and a histogram showing the integrated intensities of the blue, green, and red emission bands as a function of temperature. [Adapted with permission [153]. Copyright 2013, Royal Society of Chemistry (RSC)] **(b)** Variable-angle thermal sensing performance of the 1 mm PC@Au film@UCNPs composite, showing a schematic of the angle-

dependent measurement configuration, temperature-dependent variations of the $FIR_{520/540}$ ratio with corresponding fitted curves, and the absolute and relative sensitivities obtained at four different detection angles. [Reproduced with permission [154]. Copyright 2013, Royal Society of Chemistry (RSC)]

In an intriguing study, a 980 nm–excitable Yb/Ho/Ce:$NaGdF_4$@Yb/Tm:$NaYF_4$ core–shell nanostructure was designed to achieve simultaneous white-light emission and temperature sensing (**Figure 6a**) [153]. The spatial separation of $Ho^{3+}$ in the core and $Tm^{3+}$ in the shell suppressed deleterious energy transfer, thereby enhancing upconversion intensity, while $Ce^{3+}$ incorporation modulated the red-to-green emission ratio to realize white-light output. Temperature-dependent measurements revealed enhanced sensing performance, with a relative sensitivity of 2.4% $K^{-1}$, highlighting its promise for optical thermometry applications. In another study, a 2D PC/Au/UCNPs hybrid platform was demonstrated for high-performance temperature sensing (**Figure 6b**) [154]. Strong optical confinement arising from photonic crystal resonance and plasmonic effects significantly enhanced emission intensity and signal-to-noise ratio. The thermometric sensitivity exhibited clear angle dependence, attributed to incident direction-dependent modulation of the local electromagnetic field. However, UCNPs are commonly excited in the near-infrared region, where water exhibits relatively strong absorption that can differ across biological tissues, raising concerns about their suitability for accurate temperature sensing. In particular, the absorption coefficient of water at 980 nm is approximately 20 times higher than at 800 nm [155]. Savchuk *et al.* demonstrated that $KLuWO_4$:$Tm^{3+}$ /$Ho^{3+}$ nanoparticles function as tunable heater–thermometer nanoplatforms under 808 nm excitation [156]. The upconversion emission bands at 696 and 755 nm (within the biological window) were employed for ratiometric thermometry, yielding a maximum relative sensitivity of 2.8% $K^{-1}$ and a minimum temperature resolution of 0.2 K at 300 K. Although 800 nm excitation is often considered advantageous over 980 nm for minimizing photothermal effects in biological systems, a recent study shows that under pulsed excitation conditions suitable for single-particle imaging, neither wavelength induces significant local heating in air or aqueous environments.[157] Both 800- and 980-nm excited UCNPs demonstrate reliable ratiometric thermometry, with sensitivities of ~$10^{-4}$ $K^{-1}$. The temperature dependence follows a linear relationship between ln($I_{525}$/$I_{545}$) and inverse temperature (1/T), consistent with Boltzmann-governed population redistribution under both excitation schemes. In addition to fluorescence intensity ratios, UCNP thermometry can also utilize temperature-dependent luminescence lifetime.[158, 159] This approach generally exhibits lower sensitivity.

However, it offers advantages such as immunity to variations in nanoparticle concentration and reduced sensitivity to scattering and absorption from the surrounding medium. The lifetime variation originates from phonon-assisted nonradiative relaxation processes.

## 8. Selected Applications

### 8.1 Micro and Nano Electronics Devices

The continued miniaturization of micro- and nanoelectronic devices has rendered thermal management a fundamental bottleneck to performance, reliability, and scalability. Next-generation transistors are expected to further extend node scaling into the sub-5-nm and even sub-3-nm regimes [160]. As transistor dimensions and functional layers approach the nanometer and atomic-scale regimes, integration densities increase dramatically, while operating voltages can no longer be proportionally reduced without inducing excessive leakage currents, imposing a lower bound on device operating power [161]. This power-scaling impasse results in chip-level, highly localized thermal hotspots that accelerate device degradation and limit operational lifetimes. The challenge is further exacerbated by the rapid emergence of two-dimensional and flexible nanoelectronics. In such systems, nanoscale heat-transfer phenomena—such as suppressed thermal conductivity in ultrathin films and non-diffusive phonon transport—render classical thermal models increasingly inadequate. Although SThM and AFM-based thermocouples provide sub-100-nm spatial resolution [45, 162], their applicability is largely confined to small mapping areas and slow, contact-based measurements, which can perturb the local thermal environment. In contrast, fluorescence thermometry enables non-contact, full-field temperature mapping with sub-micrometer spatial resolution, making it particularly well suited for probing spatially extended thermal gradients and hotspot dynamics in micro- and nanoelectronic devices. In 2008, Löw *et al.* demonstrated sub-500 nm spatial-resolution surface temperature mapping using fluorescence thermometry by confining a thin layer of Rhodamine B fluorophores directly onto the surface of a resistive heating device [163]. Local Joule heating in the 80 nm-thick Ni wire led to a pronounced reduction in Rhodamine B fluorescence intensity on and near the wire, with increasingly strong fluorescence quenching observed at higher applied currents corresponding to larger temperature rises, as illustrated in **Figure 7a**.

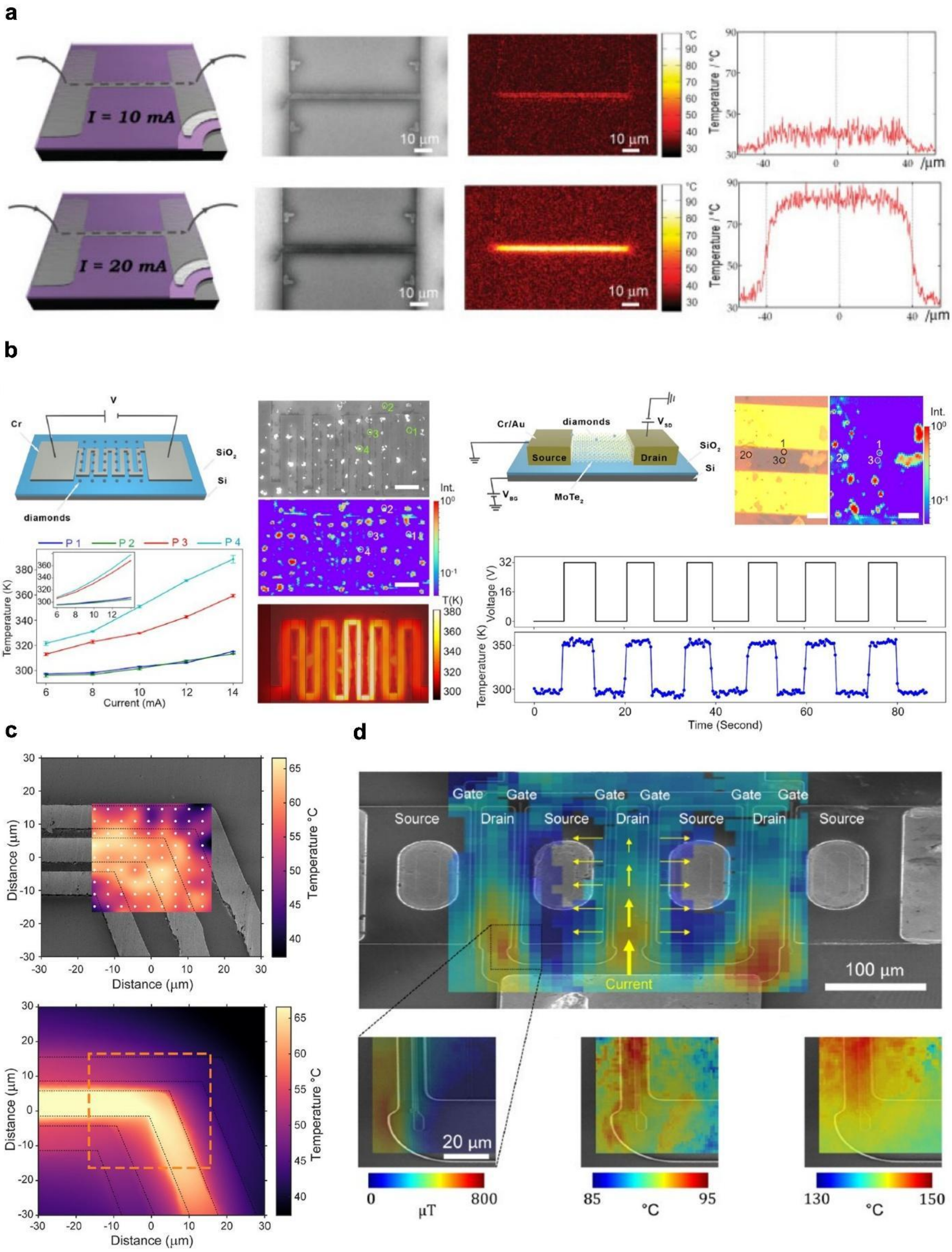

**Figure 7.** Fluorescence nanothermometry for micro/nanoelectronics. **(a)** The current-induced temperature rise was spatially mapped along an 80-µm-long, 2-µm-wide, and 40-nm-thick Ni wire using the temperature-dependent fluorescence intensity of Rhodamine B. [Reproduced with permission [163]. Copyright 2008, John Wiley and Sons] **(b)** Temperature rise measured using SiV–GeV codoped nanodiamonds at multiple locations on a microfabricated structure,

with comparison to simulations, and along an operating $MoTe_2$ flake–based FET under real-time biasing conditions. [Reproduced with permission [131]. Copyright 2023, American Chemical Society] **(c)** Experimental and simulated temperature maps of a coplanar waveguide obtained using a 9 × 9 array of nanodiamonds embedded in Polydimethylsiloxane (PDMS) platform. [Reproduced with permission [103]. Copyright 2018, American Chemical Society] **(d)** Q-CAT (quantum conformally attached thermo-magnetic) imaging of a multifinger GaN high-electron-mobility transistor (HEMT), enabling simultaneous temperature and magnetic field mapping. [Reproduced with permission [74]. Copyright 2020, American Chemical Society]

However, a key limitation of this work is that achieving a higher signal-to-noise ratio requires increased excitation power, which introduces a significant risk of photobleaching. Therefore, the excitation power must be carefully optimized to maximize signal quality while minimizing photobleaching effects. In another study [164], polymer-embedded NV-containing fluorescent nanodiamonds were employed as microscale thermometers to monitor local temperature changes in a resistively heated gold microwire, with spatial and temporal resolutions of approximately 1 μm and 0.1 s, respectively. Temperature was quantified by tracking the ZPL position and the peak-height ratio of the NV center emission in diamond. A gold microwire was embedded within a polymer film and biased with a DC voltage to induce Joule heating in the surrounding matrix, which was subsequently quantified by tracking the temperature-dependent zero-phonon line features of embedded nanodiamonds. In another study, confocal fluorescence microscopy was used to quantitatively map local temperature distributions with ~1 μm spatial resolution on a MEMS-based microheater by leveraging the ratiometric temperature dependence of the two green emission bands of $NaYF_4:Er^{3+},Yb^{3+}$ upconversion nanoparticles [165]. The microheater consists of a spiral metal element embedded in a silicon nitride membrane, which is supported by a silicon substrate. An electrical DC voltage applied across metal contacts drives a current that induces Joule heating. In another study, hexagonal boron nitride (hBN) nanothermometers were deterministically transferred onto a microcircuit using an aligned dry-transfer technique to monitor local temperature variations [166]. The microcircuit incorporated metallic contacts of varying widths—2 μm in the inner region and 4 μm in the outer region. Each hBN nanothermometer consisted of a single quantum emitter, with temperature extracted from shifts in the ZPL position. Although this approach enabled monitoring of heat dissipation at specific locations within the microcircuit, pixel-by-pixel temperature mapping across the entire device was not achievable. Chen *et al.* demonstrated real-time temperature monitoring of an operating $MoTe_2$ transistor through the use of GeV–SiV codoped nanodiamonds deposited on the $MoTe_2$ flake, as depicted by **Figure 7b** [131]. Device fabrication involved electron-beam lithography to pattern thin chromium (5 nm) and gold (80 nm) source and drain electrodes, after which a $MoTe_2$ flake was transferred to bridge the contacts. A back-gate voltage was applied to modulate the channel current. GeV–SiV codoped nanodiamonds were subsequently drop-cast onto the device to probe its operating temperature under electrical bias. When the drain–source voltage ($V_{DS}$) was increased from 16 to 24 V at a fixed back-gate voltage ($V_{BG}$) of 70 V, a temperature rise of approximately 44 K was observed. Temperature was measured by monitoring the temperature-dependent anti–Stokes–to–Stokes intensity ratio of the GeV–SiV emission. In addition to the transistor platform, the authors also fabricated a resistive microheater consisting of a 200 nm-thick chromium layer patterned by electron-beam lithography into a micrometer-scale serpentine geometry, as shown in the left panel of **Figure 7b**. Nanodiamonds were transferred onto the circuit using a prepatterned aperture array, enabling point-to-point temperature measurements. While this approach successfully demonstrated localized thermal sensing, further optimizations, such as improving nanodiamond yield within the apertures and achieving a more uniform spatial distribution, would be necessary to enable high-fidelity, pixel-resolved temperature mapping of complex integrated circuits. In another study, Andrich *et al.* fabricated dense arrays of nanodiamonds embedded within a transferable, transparent, and flexible platform using a chemical patterning technique, achieving a high device yield of 98 ± 0.8% [103]. This platform enabled spatial mapping of the temperature profile generated by a coplanar waveguide antenna, which was simultaneously used to coherently manipulate the spins of the embedded NV centers as illustrated by **Figure 7c**. The average temperature uncertainty across the array was reported to be 3.9 ± 2.9 K. However, the approach is not fully all-optical, and the use of a PDMS-based platform may introduce systematic temperature offsets due to PDMS's intrinsically low thermal conductivity, potentially affecting the accuracy of absolute temperature readout. Foy *et al.* employed ODMR-based measurements on ~100 nm-thick films of NV-containing nanodiamonds to simultaneously map magnetic fields and temperature at high frame rates (100–1000 Hz) **[74]**. When applied to a GaN HEMT, this technique revealed a pronounced temperature rise localized around the gate region, along with a sharply resolved temperature drop at the gate termination along the channel direction, as seen in **Figure 7d**. A key limitation of this work is that Q-CAT imaging relies on microwave excitation and ensemble ODMR calibration of nanodiamonds, making it not fully all-optical and limiting its applicability to devices compatible with microwave fields. In another study, hyperspectral quantum-rod thermal imaging was employed to exploit the temperature-dependent photoluminescence of CdSe/CdS quantum rods, enabling quantitative surface-temperature mapping of biased GaN HEMT devices [167]. This approach achieved a temperature precision of ~4 °C with an estimated lateral optical resolution of ~700–800 nm. A nearly uniform temperature distribution was observed in the device channel, with no detectable hotspots and only minor temperature variations.

One limitation of employing fluorescent phosphors on micro- and nanoelectronic devices is the risk of introducing unwanted surface contamination. Fiber-based thermometry offers a potential solution to this issue by enabling remote temperature sensing without direct material deposition, thereby providing a more robust and device-compatible thermal sensing approach. For example, Zhang *et al.* demonstrated a fiber-based quantum thermometer employing NV-center microdiamonds to image the surface temperature of an electronic chip [96]. Using an ODMR approach, the technique achieved a temperature sensitivity of approximately 18 mK $Hz^{-1/2}$. Results reveal a peak chip temperature of ~26.5 °C during operation, contrasted with a uniform baseline temperature of ~24.5 °C in the absence of applied power. In another work, a SiV-containing microdiamond was integrated with a multimode optical fiber to enable real-time temperature measurements of a graphite-based microheater device [129]. Instead of relying on conventional spectral analysis used in many earlier studies, this approach utilized the ratio of intensity of SiV microdiamond and backscattered laser signals, achieving an excellent temperature resolution of 0.72 K $Hz^{-1/2}$ and a spatial resolution of approximately 7.5 μm. Although the collection efficiency was relatively low due to weak coupling between the quantum material and the fiber, as well as intrinsic limitations of fiber-based collection, the method successfully captured device heating, with the temperature increasing from 298 to 318 K as the driving current was raised from 0 to 3.67 mA. This approach can be further extended to enable pixel-by-pixel thermal mapping by scanning the fiber-integrated thermal sensor across the device surface. However, such operations introduce a potential risk of mechanical detachment of the diamond sensor from the fiber, highlighting the need for more robust integration strategies.

### 8.2 Three-dimensional (3D) thermal imaging

3D temperature sensing requires the combined acquisition of thermal information and precise spatial localization. In fluorescence-based thermometry, accurate determination of the axial (z) position of the sensing material is particularly critical for reliable 3D temperature reconstruction. Fluorescence nanothermometry stands out as a highly promising approach for three-dimensional temperature sensing, delivering sub-micrometer spatial resolution (< 1 μm), high temperature sensitivity (> 1 % $K^{-1}$), excellent temperature resolution on the order of $10^{-2}$ K, and rapid response times approaching 10 μs [168]. Benninger *et al.* showed that micrometer-scale spatial resolution in three-dimensional temperature mapping across a microfluidic chip can be readily achieved using a two-photon–excitation–based optical sectioning approach [169]. In this microfluidic device configuration, the microchannel is thermally coupled to a heating element located near the upper channel surface. Fluorescence lifetime imaging of rhodamine B was performed over a range of temperatures, revealing a systematic decrease in lifetime with increasing temperature. The implementation of two-photon excitation, together with the resulting optical sectioning effect, enables a spatial resolution of ~1 μm in the lateral direction and ~10 μm along the axial dimension, as illustrated in **Figure 8a**. Owing to the long working distance requirement, a low-NA (0.25) objective was used, which limited the axial resolution and signal intensity; both can be significantly improved by employing a higher-NA objective. In 2013, Kucsko *et al.* reported the first demonstration of NV-based temperature sensing in living cells by incorporating nanodiamonds and gold nanoparticles into human embryonic fibroblast WS-1 cells [84]. Temperature was read out at the NV centers while the gold nanoparticle was locally heated, enabling intracellular temperature measurements with a sensitivity of ~80 mK $Hz^{-1/2}$. Accurate depth positioning of the nanodiamond is essential for achieving a reliable three-dimensional thermal readout. To accurately resolve the depth of NV centers beneath the diamond surface, Häußler *et al.* developed an optical approach based on axial scanning of the confocal microscope's optical axis [170]. The resulting virtual depth measurements achieved sub-nanometer precision for several NV centers. As is well established, three-dimensional temperature sensing relies on the simultaneous acquisition of both temperature and probe position. The double-helix point spread function (DH-PSF) is a method that encodes axial information into a characteristic double-helix image, enabling sub-diffraction-limit tracking of the 3D positions of micro- and nanoparticles [171]. It is an advanced microscopy approach in which the conventional point spread function, describing the image of a point emitter, is engineered into a double-helix profile by applying tailored phase masks. Here, emitters such as quantum dots appear as two lobes whose rotation angle encodes the defocus distance; analyzing this angle enables precise determination of the emitter's axial (z) position.

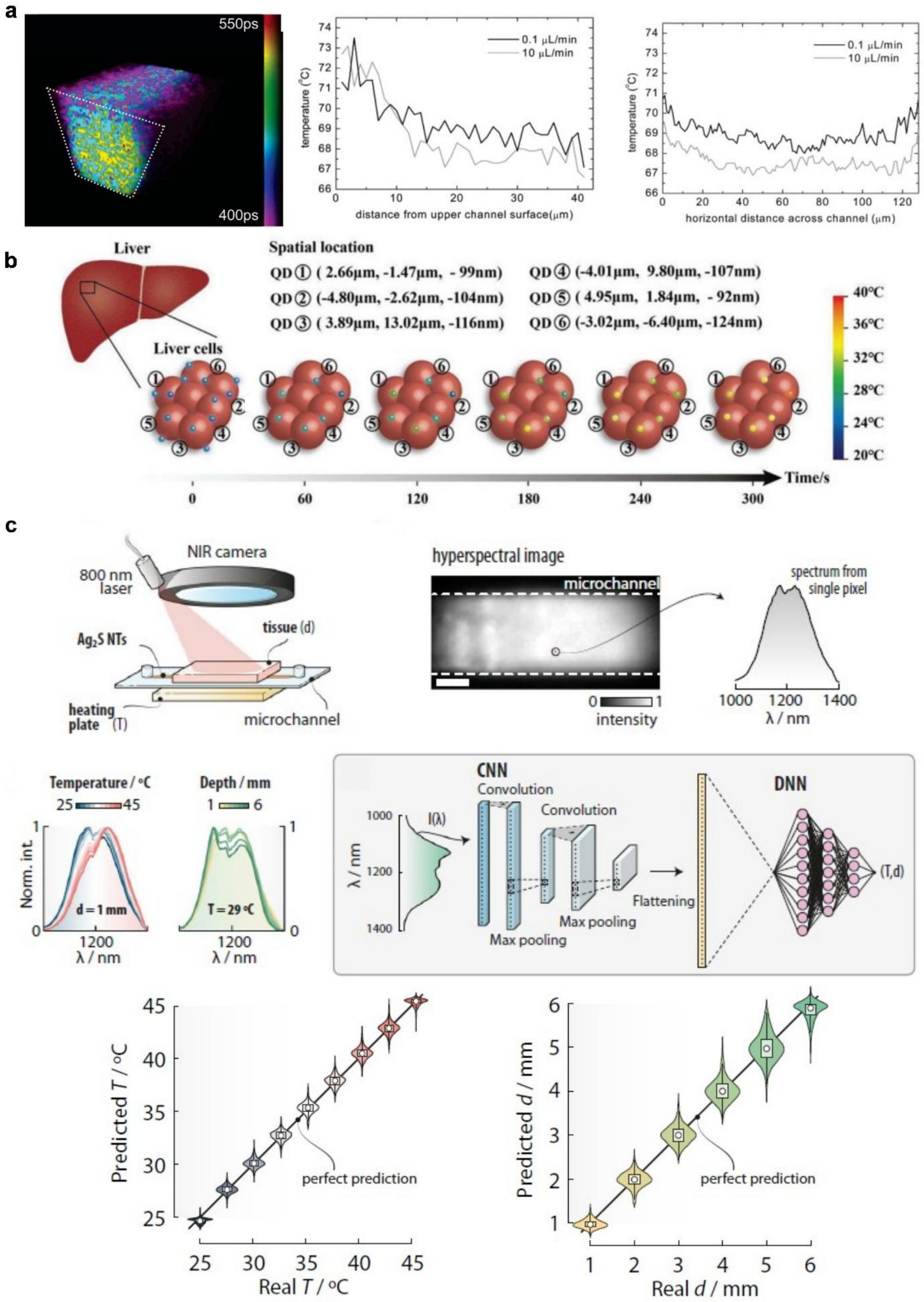

**Figure 8.** Three-dimensional temperature mapping with Fluorescence thermometry. **(a)** Rendered three-dimensional image of fluorescence lifetime data acquired from a 100-µm-long microchannel segment filled with a methanolic rhodamine B solution. Temperature profiles as a function of distance from the upper microchannel surface, together with lateral temperature distributions across the microchannel at mid-channel depth, for volumetric flow rates of 0.1 and 10 μL $min^{-1}$. [Reproduced with permission [169]. Copyright 2006, American Chemical Society] **(b)** Schematic illustration of quantum three-dimensional thermal imaging applied to liver cells. [Reproduced with permission [172]. Copyright 2009, Royal Society of Chemistry] **(c)** Calibration and machine-learning framework for three-dimensional luminescence thermometry. The figure shows the experimental setup used to calibrate fluorescent nanothermometers, representative hyperspectral images, and normalized emission spectra acquired at different temperatures and depths, the dense neural network –based algorithm employed for data analysis, and the comparison between predicted and measured temperature (T) and depth **(d)** values obtained from the trained model using both tissue phantoms and real biological tissues. [Reproduced from ref [173] under terms of the CC-BY-NC-ND 4.0. Copyright 2025, Springer Nature]

Using this approach, Yang *et al.* mapped local temperature distributions and resolved the direction of heat flux across tumor cell surfaces, employing fluorescence QDs functionalized with the membrane transport protein transferrin as targeted thermosensors [73]. The system provided an axial (z-axis) positioning range of approximately 2.6 μm for a single imaging acquisition. In another work, DH-PSF reconstruction algorithm was optimized to enhance imaging efficiency, ultimately achieving three-dimensional thermometry with CdTe quantum dots, achieving lateral and axial spatial resolutions of 130 nm and 140 nm, respectively, and a temperature resolution of 0.625 °C. The quantum dots were delivered into human tissues, including liver cells, allowing concurrent mapping of the spatial location and temperature of diseased cells, as depicted in **Figure 8b**. Different from DH-PSF–based 3D thermometry, another study achieved three-dimensional thermal imaging in mice by exploiting ratiometric emission from hybrid $Ag_2S$ QDs/$Yb^{3+}$ nanoprobes, where a temperature-insensitive $Yb^{3+}$ intensity ratio $R_1$ encodes depth, while a second ratio $R_2$ provides temperature readout under 808 nm excitation [174]. Recently, Ming *et al.* reported the first credible luminescence-enabled three-dimensional temperature imaging approach in biological tissue based on a single hyperspectral acquisition, overcoming the long-standing limitation of fluorescence thermometry to two-dimensional readouts [173]. This single-shot method combines near-infrared hyperspectral imaging of $Ag_2S$ nanothermometers with convolutional and dense neural network–based spectral decoding to simultaneously retrieve both temperature and axial depth. Because tissue-induced spectral distortions arise from a complex and highly variable interplay of absorption, scattering, composition, and photon transport that cannot be reliably captured by analytical models, a machine-learning approach is required to robustly decode temperature and depth information in fluorescence thermometry. During calibration, the emission spectra of $Ag_2S$ nanothermometers are acquired while independently and controllably varying the temperature (T) and the tissue thickness (depth, d) as illustrated in **Figure 8c.** A hybrid machine-learning framework combining a convolutional neural network (CNN) with a fully connected dense neural network (DNN) was employed, in which the CNN extracts key spectral features from the emission data and the DNN maps these features to temperature (T) and depth (d) through linear and nonlinear correlations. Results show the trained model resolves temperature and depth with precisions of approximately 0.45 °C and 0.25 mm, respectively. Although this implementation is limited by temporal and spatial resolution, this work uniquely transforms spectral distortions into a depth-encoding signal, with future improvements expected from brighter nanothermometers, more sensitive detectors, and faster hyperspectral acquisition strategies. One of the popular approach to reconstruct three-dimensional images involves acquiring a series of two-dimensional images at different axial (z) positions and stacking them to form a 3D volume, as commonly implemented in confocal laser scanning microscopy [175]. Here, the optical system transmits in-focus light while rejecting out-of-focus emission to generate a two-dimensional optical section, and three-dimensional imaging is obtained by sequentially scanning and stacking images acquired at different focal depths. Using this concept, Yu *et al.* employed rhodamine B–loaded silica ($RhB@SiO_2$) particles as thermal nanoprobes for three-dimensional temperature mapping [176]. The fluorescence intensity of the $RhB@SiO_2$ particles exhibited a linear dependence on temperature, and by axially scanning the objective lens of a confocal microscope under 785 nm excitation, optical sections and corresponding temperature maps were obtained at different z-positions.

### 8.3 Disease Diagnosis

Fluorescence nanothermometry has emerged as a powerful tool for high-resolution intracellular temperature sensing and mapping. Such capability is crucial for understanding disease-related cellular processes, including gene expression, metabolism, and cell division—which are often accompanied by subtle but measurable thermal variations[177, 178]. Core body temperature is an essential clinical indicator closely linked to disease diagnosis and therapeutic evaluation. Fever typically reflects infection or inflammation through activation of immune defenses, whereas localized cooling can result from impaired blood flow, suppressed metabolic activity, or reduced oxygen and nutrient supply [179].

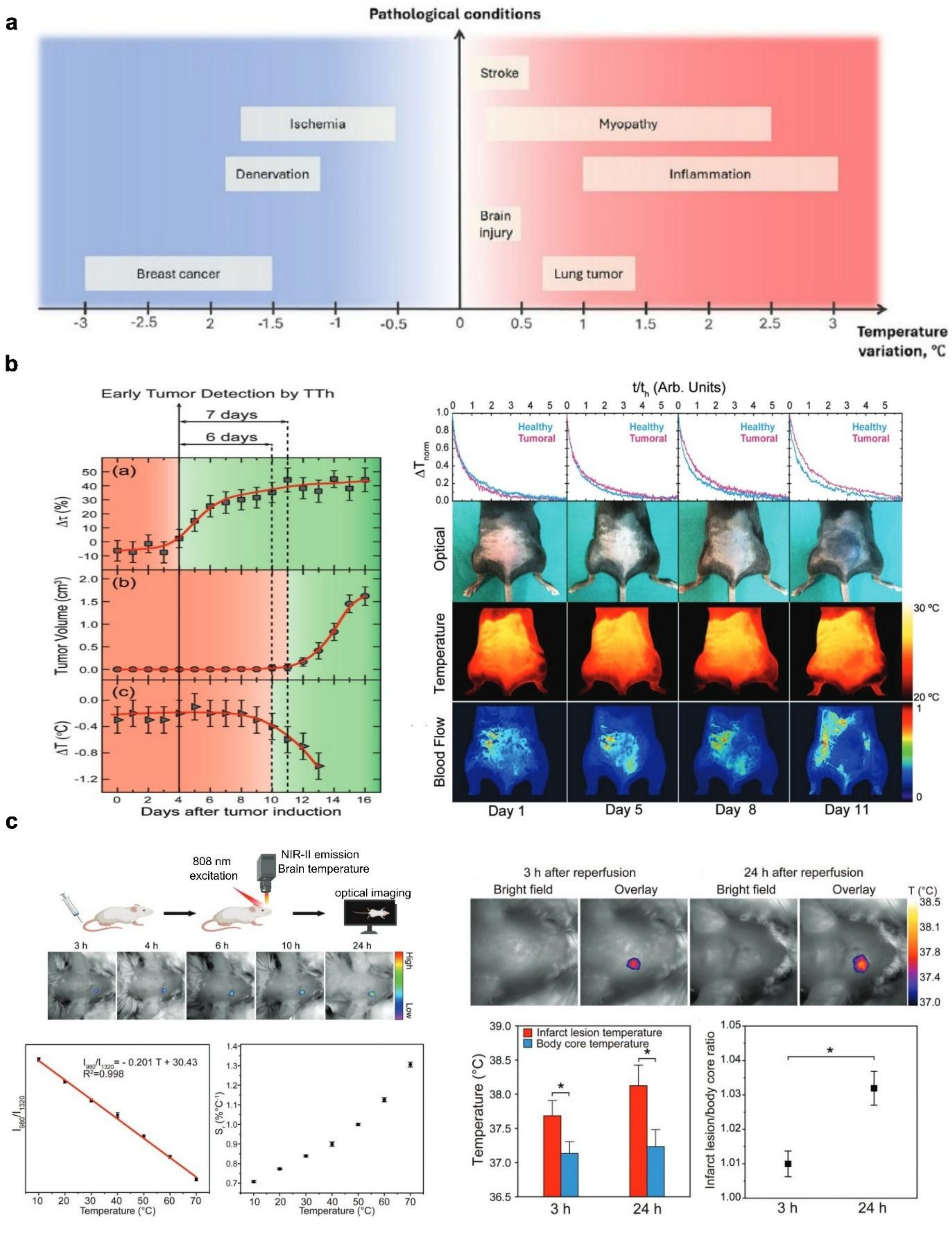

**Figure 9.** Disease diagnosis with fluorescence thermometry. **(a)** Diagram summarizing medical conditions that result in localized or systemic thermal fluctuations. [Reproduced with permission [180]. Copyright 2025, John Wiley and Sons] **(b)** Early cancer detection using fluorescence thermometry. Time-dependent evolution of $\Delta\tau$, tumor volume and the surface temperature difference between tumoral and adjacent healthy tissue following cancer cell

inoculation in a mouse model. Representative thermal transient curves of tumoral and healthy tissues recorded at different time points post-inoculation. Corresponding optical images, thermographic maps, and blood perfusion images of a representative mouse acquired at sequential stages after tumor induction. [Reproduced with permission [181]. Copyright 2018, John Wiley and Sons] **(c)** Early ischemia detection using luminescence–based thermometric imaging. Calibration curve of the luminescence intensity ratio as a function of temperature for the lanthanide-based nanothermometer. Brain temperature of the mouse model determined via NIR luminescence from the lanthanide probe. [Adapted with permission [182]. Copyright 2023, John Wiley and Sons]

The major pathological conditions associated with such temperature alterations are illustrated in **Figure 9a**. Tumors can be detected by tracking subtle temperature variations caused by changes in blood perfusion, tissue density, and metabolic activity, enabling earlier intervention and potentially preventing further cancer progression. Recently, a thermometry-based strategy using $Ag_2S$ nanoparticles was demonstrated for early in vivo tumor detection, exhibiting minimal toxicity and suitability for early-stage diagnosis [181]. $Ag_2S$ nanocrystals were injected into both tumoral and healthy tissues to monitor their thermal relaxation dynamics (**Figure 9b**, right). To achieve this, the fluorescence intensity was recorded before, during, and after moderate laser heating, allowing extraction of the transient temperature rise ($\Delta T$) from the intensity-based calibration of $Ag_2S$ emission and subsequent determination of the characteristic relaxation time ($\tau$) from the cooling curve. On day 1, the relaxation curves of healthy and tumoral tissues were nearly identical. By day 5, although the tumor remained undetectable by visual inspection or infrared thermography, significant differences in the thermal relaxation times enabled its identification. These disparities became progressively more pronounced by day 11, clearly distinguishing tumoral from healthy tissue. The diagnostic capability of this approach is further highlighted in **Figure 9b** (left), which shows the temporal evolution of the transient temperature increase following tumor induction, along with corresponding changes in tumor volume and the surface temperature difference measured by infrared thermography. In another study, $Nd^{3+}$-doped $LaF_3$ nanocrystals were employed to regulate photothermal therapy, simultaneously enabling optical heating and temperature monitoring [183]. Local temperature elevation is closely linked to inflammatory responses. In this context, a mouse inflammation model was employed to assess the in vivo performance of hybrid nanothermometers comprising temperature-sensitive triplet–triplet annihilation upconversion units integrated with temperature-insensitive Nd-doped nanoparticles.[184]. These hybrid ratiometric nanothermometers enabled precise in vivo temperature monitoring, achieving high thermal sensitivity (~7.1% $K^{-1}$) and excellent resolution (~0.1 K). Nevertheless, their practical utility is constrained by significant attenuation of visible light in biological tissues, resulting in reduced signal-to-noise ratios and the need for dual-beam excitation, which increases experimental complexity. Li *et al.* introduced a temperature-sensitive, near-infrared emitting lanthanide-based nanothermometer designed for early ischemic stroke detection and real-time monitoring of in vivo brain temperature [182]. Using luminescence intensity ratio ($I_{980}/I_{1320}$) thermometry, intracerebral temperature elevation in the infarct area was quantitatively monitored in a mouse model as depicted in **Figure 9c.** The results showed that the infarcted brain regions exhibited significantly higher temperatures than the core body temperature, with a progressive increase observed from 3 to 24 h after reperfusion. Moreover, the temperature ratio (ischemic brain/body core) increased over time, indicating that elevated brain temperature correlates with the progression of the pathological condition, as shown in **Figure 9c** (right bottom). Compared with conventional medical imaging modalities such as computed tomography and magnetic resonance imaging, this approach offers a more cost-effective, portable, and rapid bedside diagnostic alternative, enabling earlier and more accessible detection of ischemic stroke. In another study, PbS/CdS/ZnS quantum dots were utilized as deep-tissue nanothermometers to monitor ischemia in mice and to distinguish between ischemic and inflammatory phases after femoral artery double ligation [185]. In 2024, a ~1550 nm–emitting lanthanide nanothermometer enabled minimally invasive deep-brain vascular thermal imaging in live mice, achieving ~200 μm spatial resolution [186]. For biomedical applications, the phototoxicity of phosphors must be carefully addressed. For instance, $Ag_2S$ nanoparticles generally exhibit lower phototoxicity than UCNPs because they operate in the NIR-II window, where water and biological tissues exhibit minimal absorption, thereby reducing photothermal heating and reactive oxygen species generation[187]. In contrast, UCNPs excited at 980 nm can induce local heating through water absorption, leading to moderate photothermal effects; however, this can be mitigated by employing 808 nm excitation or optimized nanostructures.

## 9. Conclusion and Outlook

The wide range of fluorescent nanothermometers highlighted in this review underscores the growing influence of nanothermometry across diverse disciplines. Current applications extend from micro- and nanoelectronic systems to nanofluidics, disease diagnostics, and three-dimensional thermal imaging in biological environments. Although the field has advanced rapidly over the past two decades in response to increasing technological demands, important challenges remain.

### 9.1 Challenges

Although fluorescence nanothermometry is a well-established and versatile research field with wide-ranging applications, its translation into commercial devices remains limited. Variations in nanomaterial quality from batch to batch necessitate individual sensor calibration, which can be labor-intensive and impractical for large-scale applications. Furthermore, investigating heat generation in miniaturized electronic devices or thermal behavior within biological microenvironments requires high spatial resolution. However, in fluorescence thermometry, the achievable spatial resolution is fundamentally limited by optical diffraction, which is governed by the numerical aperture (NA) of the imaging system—a measure of the objective lens's light-gathering capability. According to the Rayleigh–Abbe criterion, the minimum resolvable distance is approximately $\lambda/(2NA)$ [188]. For visible wavelengths (~500 nm) and a high-NA objective (NA ≈ 1.4), this results in a diffraction limit of roughly 200 nm. Therefore, features spaced closer than this distance cannot be resolved independently and appear blurred in the resulting image. The ultimate limit of spatial resolution in fluorescence thermometry is achieved in single-nanoparticle sensing. Although both excitation and emission remain diffraction-limited in the far field, the effective spatial resolution is governed solely by the physical size of the nanothermometer, provided that the detected signal originates exclusively from that individual nanoparticle. A further challenge in fluorescence thermometry is achieving reliable pixel-to-pixel thermal mapping. Variations in nanoparticle distribution, emission intensity, and local calibration factors across the imaging field can introduce spatial inconsistencies, limiting quantitative temperature accuracy at the single-pixel level. A key parameter in assessing the performance of a luminescent thermometer is the signal-to-noise ratio (SNR). When analyzing only a subset of pixels in a fluorescence or hyperspectral image, detector performance becomes critical, as the SNR is governed by the number of photons collected per pixel within the chosen exposure time. In biological systems, a uniform distribution of nanothermometers is rarely achievable; instead, targeting strategies are typically used to promote selective accumulation at specific regions of interest, such as particular cellular compartments. Under these conditions, selecting and analyzing specific pixels within the luminescence or hyperspectral image remains valid and often necessary for accurate temperature readout [189]. The intrinsic brightness of a nanothermometer plays a critical role in determining the signal-to-noise ratio. High brightness allows efficient signal detection under low excitation irradiance, thereby reducing photodamage and minimizing laser-induced self-heating. A high quantum yield further ensures that a substantial fraction of absorbed photons is re-emitted as fluorescence, improving sensitivity and measurement reliability. However, a major limitation of most fluorescent materials is thermal quenching, in which the emission intensity decreases at elevated temperatures, compromising performance. Encouragingly, recent advances have identified materials exhibiting positive temperature-dependent emission, offering a promising strategy to overcome this constraint and enhance thermometric robustness [190].

Biased temperature readout represents a critical artifact in fluorescent nanothermometry and must be carefully addressed to ensure measurement reliability. One major source of error arises from excitation power–induced bias. Any discrepancy between the excitation power density used during calibration and that employed during actual measurements can result in substantial temperature inaccuracies. This problem is particularly pronounced in ratiometric thermometry when the two emission bands originate from different excitation mechanisms—for instance, a linear one-photon process ($I_1 \propto P$) and a nonlinear two-photon or upconversion process ($I_2 \propto P^2$) [15]. Under such conditions, the fluorescence intensity ratio ($I_2/I_1$) becomes intrinsically dependent on excitation power, introducing systematic errors unless strict power stabilization and calibration consistency are maintained. Biased temperature readings can also result from laser-induced self-heating, especially when nanoparticles are deposited on substrates with low thermal conductivity or when surface contamination is present [135]. Under such conditions, even moderate excitation powers may produce significant localized heating, thereby affecting both calibration and actual temperature measurements. Substrates with poor thermal conductivity are particularly susceptible to this effect. In addition, polymer interlayers on otherwise thermally conductive substrates, such as silicon, can impede heat dissipation, leading to laser-induced temperature increases with increasing polymer thickness. These factors can generate notable temperature offsets at sub-micrometer and even nanometer scales and therefore must be carefully considered in nanoscale thermometry studies. In certain applications, luminescent thermometers are deposited directly onto surfaces, such as microelectronic devices. In these configurations, part of the emitted fluorescence may be reflected by the underlying surface, modifying the detected signal and potentially biasing the temperature readout [191]. This reflection can be either constructive or destructive, depending on factors such as the separation between the thermometer and the metallic surface, the emission wavelength, and the angle of emission or reflection. As a result of these complex optical interactions, the spectral profile collected by the detection system may differ from the nanothermometer's intrinsic emission.

### 9.2 Future Directions

Only a limited number of studies have systematically addressed particle-to-particle variability, even though such differences can represent a major, sometimes dominant, source of uncertainty when using uncalibrated nanothermometers. In single-particle measurements, calibration across multiple individual particles should therefore be reported to ensure reliability. Similar considerations apply to ensemble measurements, where calibration at different spatial locations may be necessary when particles are dispersed on substrates. In temperature mapping applications, achieving the highest accuracy may even require pixel-by-pixel calibration. These practical limitations motivate the development of alternative strategies to reduce calibration burden. Multiparametric approaches in luminescence thermometry provide one such solution. By simultaneously leveraging multiple temperature-dependent optical features, such as spectral position, linewidth, intensity, lifetime, or intensity ratios, these methods reduce reliance on strict single-parameter calibration for each nanoparticle. Analytical frameworks, including multiple linear regression, dimensionality reduction, and neural networks, integrate diverse spectral descriptors to produce robust temperature predictions. This significantly decreases calibration effort while improving precision and lowering temperature uncertainty compared to conventional single-feature methods. Choi *et al.* introduced a multiparametric strategy that integrates several PL features to accurately determine temperature [127]. By combining multiple spectral parameters rather than relying on a single observable, they achieved a high thermal resolution of 13 mK $Hz^{-1/2}$, significantly outperforming conventional single-parameter approaches. In another study, a machine-learning-based multifeature regression model was implemented that required calibration at only three temperature points using five training nanodiamonds, yet still enabled reliable, accurate temperature prediction for unseen nanodiamonds [192]. Maturi *et al.* demonstrated a tenfold improvement in multiparametric nanothermometry over traditional single-feature methods, achieving a record relative sensitivity of 50% $K^{-1}$ and a temperature uncertainty below 0.1 K [193]. In a recent study, a two-stage machine learning–based multi-feature regression model integrating both temperature-dependent Raman and PL features achieved temperature accuracies as low as 0.7 K, resolutions down to 0.6 $K \cdot Hz^{-1/2}$, and sensitivities as high as 0.04 $K^{-1}$ [194]. These results demonstrate that multiparametric strategies combined with machine learning represent a major step forward for fluorescent nanothermometry, enhancing performance and facilitating broader adoption across diverse scientific applications. In vivo temperature fluctuations in biological systems demand a thermal resolution better than 0.1 °C and a temporal resolution of ~0.1 s to accurately capture rapid thermal dynamics, enabling broader biomedical applications. Furthermore, achieving true 3D thermal imaging requires accurate depth determination of individual nanothermometers, along with real-time collection of their temperature-dependent spectroscopic responses. Conversely, since spectral distortions vary with tissue thickness, analyzing the emitted spectra can also provide information about probe depth. A comparable concept has been demonstrated using thermal radiation to retrieve depth-dependent temperature profiles [195]. By integrating these strategies, it may become feasible to construct comprehensive 3D functional thermal maps, underscoring the importance of nanothermometers with high lifetime-based thermal sensitivity and broad emission bandwidths. Identifying micro- and nanoscale hotspots in electronic devices, among the earliest device-driven applications of luminescence thermometry, remains an important research frontier. The rise of advanced platforms, particularly wide-bandgap semiconductors, has introduced new thermal management demands requiring improved metrology tools. This need is further amplified by the development of 2D material-based devices, in which heat generation is confined to a few atomic layers, making accurate nanoscale thermal characterization even more critical.

Developing nanothermometers capable of simultaneously probing additional thermodynamic parameters, such as pressure, electric field, or chemical environment, represents an important future direction toward multimodal nanosensing platforms [196]. Such multimodal platforms would enable a more comprehensive understanding of coupled processes, particularly in complex systems like living cells and nanoelectronic devices, where thermal, electrical, and mechanical effects are inherently interdependent. In this context, $YF_3:Yb^{3+}$–$Er^{3+}$ upconverting microparticles act as bifunctional luminescent sensors for simultaneous temperature and pressure measurement, enabled by nearly independent spectroscopic parameters of Erbium ($Er^{3+}$) emission [197]. The luminescent intensity ratio is highly temperature-sensitive but pressure-insensitive, whereas the emission peak positions are weakly temperature-dependent but strongly pressure-sensitive, allowing decoupled determination of both parameters. The integration of multimodal sensors into high-resolution 2D and 3D imaging platforms presents a significant opportunity for the field. Achieving real-time, spatially resolved mapping of both temperature and pressure would greatly enhance capabilities across a wide range of applications, including aerospace systems, geothermal monitoring, live biological imaging, and the investigation of mechanical stress in functional materials.

In summary, fluorescence nanothermometry has emerged as a powerful and versatile sensing approach with substantial potential for future advancement. Addressing current limitations will require close interdisciplinary collaboration across chemistry, physics, engineering, and life sciences. Such convergence of expertise is expected to accelerate innovation, enabling the development of more robust, accurate, and application-ready platforms.

**Acknowledgements**

((Acknowledgements, general annotations))

**Data Availability Statement**

((include as appropriate, including link to repository))

## *References*

1. Wang, C.; Xu, R.; Tian, W., et al., *Cell Research* **2011,** *21* (10), 1517–1519. DOI 10.1038/cr.2011.117.
2. Okabe, K.; Uchiyama, S., *Communications Biology* **2021,** *4* (1), 1377. DOI 10.1038/s42003-021-02908-2.
3. Rosso, L.; Tabandeh, S.; Beltramino, G.; Fernicola, V., *Measurement Science and Technology* **2020,** *31* (3), 034002. DOI 10.1088/1361-6501/ab4b6b.
4. Swoboda, T.; Gao, X.; Rosário, C. M. M., et al., *ACS Applied Electronic Materials* **2023,** *5* (9), 5025–5031. DOI 10.1021/acsaelm.3c00782.
5. Li, G.; Su, Z.; Li, M., et al., *Advanced Energy Materials* **2022,** *12* (48), 2202887. DOI https://doi.org/10.1002/aenm.202202887.
6. Li, F.; Xue, H.; Lin, X.; Zhao, H.; Zhang, T., *Acs Applied Materials & Interfaces* **2022,** *14* (38), 43844–43852. DOI 10.1021/acsami.2c15687.
7. Jurga, N.; Runowski, M.; Grzyb, T., *Journal of Materials Chemistry C* **2024,** *12* (32), 12218–12248. DOI 10.1039/d3tc04716d.
8. Ross-Pinnock, D.; Maropoulos, P. G., *Proceedings of the Institution of Mechanical Engineers, Part B: Journal of Engineering Manufacture* **2016,** *230* (5), 793–806. DOI 10.1177/0954405414567929.
9. Temperature Sensors Market Size & Share Analysis https://www.mordorintelligence.com/industry-reports/temperature-sensors-market-industry (accessed March 18, 2026).
10. Allison, S. W.; Gillies, G. T., *Review of Scientific Instruments* **1997,** *68* (7), 2615–2650. DOI 10.1063/1.1148174.
11. Hoogeboom-Pot, K. M.; Hernandez-Charpak, J. N.; Gu, X., et al., *Proceedings of the National Academy of Sciences* **2015,** *112* (16), 4846–4851. DOI 10.1073/pnas.1503449112.
12. Brites, C. D. S.; Balabhadra, S.; Carlos, L. D., *Advanced Optical Materials* **2019,** *7* (5), 1801239. DOI https://doi.org/10.1002/adom.201801239.

13. Brites, C. D.; Lima, P. P.; Silva, N. J., et al., *Nanoscale* **2012,** *4* (16), 4799–829. DOI 10.1039/c2nr30663h.
14. Brites, C. D. S.; Marin, R.; Suta, M., et al., *Advanced Materials* **2023,** *35* (36), 2302749. DOI https://doi.org/10.1002/adma.202302749.
15. Zhou, J.; del Rosal, B.; Jaque, D.; Uchiyama, S.; Jin, D., *Nature Methods* **2020,** *17* (10), 967–980. DOI 10.1038/s41592-020-0957-y.
16. Harrington, B.; Ye, Z.; Signor, L.; Pickel, A. D., *ACS Nanoscience Au* **2024,** *4* (1), 30–61. DOI 10.1021/acsnanoscienceau.3c00051.
17. Bai, T.; Gu, N., *Small* **2016,** *12* (34), 4590–4610.
18. del Rosal, B.; Ximendes, E.; Rocha, U.; Jaque, D., *Advanced Optical Materials* **2017,** *5* (1), 1600508.
19. Jaque, D.; Vetrone, F., *Nanoscale* **2012,** *4* (15), 4301–4326.
20. Quintanilla, M.; Henriksen-Lacey, M.; Renero-Lecuna, C.; Liz-Marzán, L. M., *Chemical Society Reviews* **2022,** *51* (11), 4223–4242. DOI 10.1039/D2CS00069E.
21. Nexha, A.; Carvajal, J. J.; Pujol, M. C.; Díaz, F.; Aguiló, M., *Nanoscale* **2021,** *13* (17), 7913–7987. DOI 10.1039/d0nr09150b.
22. Rodríguez-Sevilla, P.; Marin, R.; Ximendes, E., et al., *Frontiers in Chemistry* **2022,** *Volume 10 - 2022*. DOI 10.3389/fchem.2022.941861.
23. Suo, H.; Zhao, X.; Zhang, Z., et al., *Laser & Photonics Reviews* **2021,** *15* (1), 2000319. DOI https://doi.org/10.1002/lpor.202000319.
24. Michalski, L.; Eckersdorf, K.; Kucharski, J.; McGhee, J., *Measurement Science and Technology* **2002,** *13* (10), 1651. DOI 10.1088/0957-0233/13/10/702.
25. Bradac, C.; Lim, S. F.; Chang, H. C.; Aharonovich, I., *Advanced Optical Materials* **2020,** *8* (15), 2000183. DOI 10.1002/adom.202000183.
26. Steur, P. P. M.; Durieux, M., *Metrologia* **1986,** *23* (1), 1. DOI 10.1088/0026-1394/23/1/002.
27. Qu, J. F.; Benz, S. P.; Rogalla, H., et al., *Meas Sci Technol* **2019,** *30* (11). DOI 10.1088/1361-6501/ab3526.
28. Spietz, L.; Lehnert, K. W.; Siddiqi, I.; Schoelkopf, R. J., *Science* **2003,** *300* (5627), 1929–32. DOI 10.1126/science.1084647.
29. Bell, J. F. W., *Ultrasonics* **1968,** *6* (1), 11–14. DOI https://doi.org/10.1016/0041-624X(68)90009-7.
30. Blanchet, F.; Chang, Y.-C.; Karimi, B.; Peltonen, J. T.; Pekola, J. P., *Physical Review Applied* **2022,** *17* (1), L011003. DOI 10.1103/PhysRevApplied.17.L011003.
31. Fellmuth, B.; Fischer, J.; Machin, G., et al., *Philos Trans A Math Phys Eng Sci* **2016,** *374* (2064), 20150037. DOI 10.1098/rsta.2015.0037.
32. Rubin, L. G., *Cryogenics* **1997,** *37* (7), 341–356. DOI https://doi.org/10.1016/S0011-2275(97)00009-X.
33. Rusby, R.; Hudson, R.; Durieux, M., et al., *Metrologia* **1991,** *28* (1), 9.
34. Carlos, L. D.; Palacio, F., *Thermometry at the Nanoscale: Techniques and Selected Applications*. RSC: **2015**.
35. Optical Temperature Sensors Market Size, Future Growth and Forecast 2033. https://www.strategicrevenueinsights.com/industry/optical-temperature-sensors-market (accessed March 18, 2025).
36. Zhang, Y.; Zhu, W.; Hui, F., et al., *Advanced Functional Materials* **2020,** *30* (18), 1900892. DOI https://doi.org/10.1002/adfm.201900892.
37. Palmer, L. D.; Lee, W.; Durham, D. B., et al., *ACS Physical Chemistry Au* **2025,** *5* (6), 589–598. DOI 10.1021/acsphyschemau.5c00044.
38. Meola, C., *Infrared Thermography: Recent Advances and Future Trends* **2012**, 3–28. DOI 10.2174/978160805143411201010003.

39. Mohan, R.; Khan, S.; Wilson, R. B.; Hopkins, P. E., *Nature Reviews Methods Primers* **2025,** *5* (1), 55. DOI 10.1038/s43586-025-00425-8.
40. Cahill, D. G., *Review of Scientific Instruments* **2004,** *75* (12), 5119–5122. DOI 10.1063/1.1819431.
41. Cahill, D. G.; Goodson, K.; Majumdar, A., *Journal of Heat Transfer* **2001,** *124* (2), 223–241. DOI 10.1115/1.1454111.
42. Tuschel, D.; Adar, F., *Spectroscopy* **2016,** *31* (12), 8–13.
43. Meng, Q.; Zhang, J.; Zhang, Y., et al., *Science Advances* **2024,** *10* (3), eadl1015. DOI doi:10.1126/sciadv.adl1015.
44. Gomès, S.; Assy, A.; Chapuis, P.-O., *physica status solidi (a)* **2015,** *212* (3), 477–494. DOI 10.1002/pssa.201400360.
45. Sadat, S.; Tan, A.; Chua, Y. J.; Reddy, P., *Nano Letters* **2010,** *10* (7), 2613–2617. DOI 10.1021/nl101354e.
46. Prasher, R. S.; Phelan, P. E., *Journal of Applied Physics* **2006,** *100* (6). DOI 10.1063/1.2353704.
47. Mecklenburg, M.; Hubbard, W. A.; White, E. R., et al., *Science* **2015,** *347* (6222), 629–632. DOI 10.1126/science.aaa2433.
48. Deng, B.; Wu, W.; Li, X., et al., *IEEE Transactions on Industrial Electronics* **2022,** *69* (11), 11774–11784. DOI 10.1109/tie.2021.3120471.
49. Landmann, M.; Heist, S.; Dietrich, P., et al., *Optics and Lasers in Engineering* **2019,** *121*, 448–455. DOI 10.1016/j.optlaseng.2019.05.009.
50. Cerruti, M. G.; Sauthier, M.; Leonard, D., et al., *Analytical Chemistry* **2006,** *78* (10), 3282–3288. DOI 10.1021/ac0600555.
51. Jiang, P.; Qian, X.; Yang, R., *Journal of Applied Physics* **2018,** *124* (16), 161103. DOI 10.1063/1.5046944.
52. Milich, M.; Olson, D. H.; Tiernan, E. M., et al., *Acta Materialia* **2025,** *288*, 120802. DOI 10.1016/j.actamat.2025.120802.
53. Koh, Y. K.; Bae, M.-H.; Cahill, D. G.; Pop, E., *Nano Letters* **2010,** *10* (11), 4363–4368. DOI 10.1021/nl101790k.
54. Ngo, D. N.; Ho, V. T. T. X.; Kim, G., et al., *Analytical Chemistry* **2022,** *94* (17), 6463–6472. DOI 10.1021/acs.analchem.1c04452.
55. Oudjedi, F.; Lee, S. S.; Paliouras, M., et al., *ACS Applied Nano Materials* **2024,** *7* (17), 20942–20953. DOI 10.1021/acsanm.4c03865.
56. Park, T.; Guan, Y.-J.; Liu, Z.-Q.; Zhang, Y., *Physical Review Applied* **2018,** *10* (3), 034049. DOI 10.1103/PhysRevApplied.10.034049.
57. Kumar, K.; Stefanczyk, O.; Chorazy, S.; Nakabayashi, K.; Ohkoshi, S.-i., *Advanced Optical Materials* **2022,** *10* (22), 2201675. DOI https://doi.org/10.1002/adom.202201675.
58. Li, P.; Askes, S. H. C.; del Pino Rosendo, E., et al., *The Journal of Physical Chemistry C* **2023,** *127* (20), 9690–9698. DOI 10.1021/acs.jpcc.3c01561.
59. Rodríguez-Sevilla, P.; Spicer, G.; Sagrera, A., et al., *Advanced Optical Materials* **2023,** *11* (11), 2201664. DOI https://doi.org/10.1002/adom.202201664.
60. Tovee, P. D.; Kolosov, O. V., *Nanotechnology* **2013,** *24* (46), 465706. DOI 10.1088/0957-4484/24/46/465706.
61. Nguyen, K. L.; Merchiers, O.; Chapuis, P. O., *Journal of Quantitative Spectroscopy and Radiative Transfer* **2017,** *202*, 154–167. DOI https://doi.org/10.1016/j.jqsrt.2017.07.021.
62. Shen, L.; Mecklenburg, M.; Dhall, R.; Regan, B. C.; Cronin, S. B., *Applied Physics Letters* **2019,** *115* (15). DOI 10.1063/1.5094443.
63. Hu, X.; Yasaei, P.; Jokisaari, J., et al., *Physical Review Letters* **2018,** *120* (5), 055902. DOI 10.1103/PhysRevLett.120.055902.
64. Dramićanin, M. D., *Methods Appl Fluoresc* **2016,** *4* (4), 042001. DOI 10.1088/2050-6120/4/4/042001.

65. Neubert, P. Radiation thermometer. **1937**.
66. Allison, S. W., *Measurement Science and Technology* **2019,** *30* (7), 072001. DOI 10.1088/1361-6501/ab1d02.
67. Urbach, F.; Nail, N. R.; Pearlman, D., *Journal of the Optical Society of America* **1949,** *39* (12), 1011–1019. DOI 10.1364/JOSA.39.001011.
68. Advance Energy Industries. https://www.advancedenergy.com/en-us/products/sense-and-measurement/thermal-sensing/fiber-optic-sensors/ (accessed 30 January).
69. Wang, S.; Westcott, S.; Chen, W., *The Journal of Physical Chemistry B* **2002,** *106* (43), 11203–11209. DOI 10.1021/jp026445m.
70. Uchiyama, S.; Matsumura, Y.; de Silva, A. P.; Iwai, K., *Analytical Chemistry* **2003,** *75* (21), 5926–5935. DOI 10.1021/ac0346914.
71. Brites, C. D. S.; Marin, R.; Suta, M., et al., *Advanced Materials* **2023,** *35*, 2302749. DOI 10.1002/adma.202302749.
72. del Rosal, B.; Carrasco, E.; Ren, F., et al., *Advanced Functional Materials* **2016,** *26* (33), 6060–6068. DOI https://doi.org/10.1002/adfm.201601953.
73. Yang, J.; Du, H.; Chai, Z., et al., *Small* **2021,** *17* (39), 2102807. DOI https://doi.org/10.1002/smll.202102807.
74. Foy, C.; Zhang, L.; Trusheim, M. E., et al., *Acs Applied Materials & Interfaces* **2020,** *12* (23), 26525–26533. DOI 10.1021/acsami.0c01545.
75. Vogel, R.; Groefsema, D. W.; van den Bulk, M. A., et al., *Acs Applied Materials & Interfaces* **2025,** *17* (14), 21215–21222. DOI 10.1021/acsami.5c00243.
76. Maturi, F. E.; Brites, C. D. S.; Silva, R. R., et al., *Advanced Photonics Research* **2022,** *3* (6), 2100227. DOI https://doi.org/10.1002/adpr.202100227.
77. Brites, C. D. S.; Xie, X.; Debasu, M. L., et al., *Nature Nanotechnology* **2016,** *11* (10), 851–856. DOI 10.1038/nnano.2016.111.
78. Aharonovich, I.; Greentree, A. D.; Prawer, S., *Nature Photonics* **2011,** *5* (7), 397–405. DOI 10.1038/nphoton.2011.54.
79. Aharonovich, I.; Neu, E., *Advanced Optical Materials* **2014,** *2* (10), 911–928. DOI https://doi.org/10.1002/adom.201400189.
80. Doherty, M. W.; Manson, N. B.; Delaney, P., et al., *Physics Reports* **2013,** *528* (1), 1–45. DOI https://doi.org/10.1016/j.physrep.2013.02.001.
81. Belser, S.; Hart, J.; Gu, Q.; Shanahan, L.; Knowles, H. S., *Applied Physics Letters* **2023,** *123* (2). DOI 10.1063/5.0147469.
82. Cambria, M. C.; Thiering, G.; Norambuena, A., et al., *Physical Review B* **2023,** *108* (18), L180102. DOI 10.1103/PhysRevB.108.L180102.
83. Lee, Y.; Kim, K.; Kim, D.; Lee, J. S., *Journal of the American Chemical Society* **2025,** *147* (16), 13180–13189. DOI 10.1021/jacs.4c16365.
84. Kucsko, G.; Maurer, P. C.; Yao, N. Y., et al., *Nature* **2013,** *500* (7460), 54–58. DOI 10.1038/nature12373.
85. Neumann, P.; Jakobi, I.; Dolde, F., et al., *Nano Letters* **2013,** *13* (6), 2738–2742. DOI 10.1021/nl401216y.
86. Toyli, D. M.; de las Casas, C. F.; Christle, D. J.; Dobrovitski, V. V.; Awschalom, D. D., *Proc Natl Acad Sci U S A* **2013,** *110* (21), 8417–21. DOI 10.1073/pnas.1306825110.
87. Acosta, V. M.; Bauch, E.; Ledbetter, M. P., et al., *Physical Review Letters* **2010,** *104* (7), 070801. DOI 10.1103/PhysRevLett.104.070801.
88. Fujiwara, M.; Shikano, Y., *Nanotechnology* **2021,** *32* (48). DOI 10.1088/1361-6528/ac1fb1.
89. Tetienne, J. P.; Hingant, T.; Rondin, L., et al., *Physical Review B* **2013,** *87* (23), 235436. DOI 10.1103/PhysRevB.87.235436.
90. Tzeng, Y.-K.; Tsai, P.-C.; Liu, H.-Y., et al., *Nano Letters* **2015,** *15* (6), 3945–3952. DOI 10.1021/acs.nanolett.5b00836.

91. Wang, N.; Liu, G.-Q.; Leong, W.-H., et al., *Physical Review X* **2018,** *8* (1), 011042. DOI 10.1103/PhysRevX.8.011042.
92. Fujiwara, M.; Sun, S.; Dohms, A., et al., *Science Advances* **2020,** *6* (37), eaba9636. DOI doi:10.1126/sciadv.aba9636.
93. Wang, J.; Feng, F.; Zhang, J., et al., *Physical Review B* **2015,** *91* (15), 155404. DOI 10.1103/PhysRevB.91.155404.
94. Liu, G.-Q.; Feng, X.; Wang, N.; Li, Q.; Liu, R.-B., *Nature communications* **2019,** *10* (1), 1344. DOI 10.1038/s41467-019-09327-2.
95. Abrahams, G. J.; Ellul, E.; Robertson, I. O., et al., *Physical Review Applied* **2023,** *19* (5), 054076. DOI 10.1103/PhysRevApplied.19.054076.
96. Zhang, S.-C.; Dong, Y.; Du, B., et al., *Review of Scientific Instruments* **2021,** *92* (4), 044904. DOI 10.1063/5.0044824.
97. Chen, G.; Wu, D.; Xue, Y., et al., *Review of Scientific Instruments* **2023,** *94* (10). DOI 10.1063/5.0146076.
98. Li, P.; Li, X.; Wang, X., et al., *Advanced Quantum Technologies* **2025,** *8* (9), 2400664. DOI https://doi.org/10.1002/qute.202400664.
99. Fedotov, I. V.; Blakley, S.; Serebryannikov, E. E., et al., *Applied Physics Letters* **2014,** *105* (26), 261109. DOI 10.1063/1.4904798.
100. Zhang, T.; Liu, G. Q.; Leong, W. H., et al., *Nat Commun* **2018,** *9* (1), 3188. DOI 10.1038/s41467-018-05673-9.
101. Yang, H.; Chen, G. B.; Zhao, X. T.; He, F. Y.; Du, G. X., *IEEE Sensors Journal* **2023,** *23* (23), 28633–28639. DOI 10.1109/JSEN.2023.3325832.
102. Gu, Q.; Shanahan, L.; Hart, J. W., et al., *ACS Nano* **2023,** *17* (20), 20034–20042. DOI 10.1021/acsnano.3c05285.
103. Andrich, P.; Li, J.; Liu, X., et al., *Nano Letters* **2018,** *18* (8), 4684–4690. DOI 10.1021/acs.nanolett.8b00895.
104. Wu, K.; Lu, Q.; Ren, Y., et al., *Advanced Materials n/a* (n/a), e17076. DOI https://doi.org/10.1002/adma.202517076.
105. Gupta, M.; Zhang, T.; Yeung, L., et al., *Advanced Sensor Research* **2024,** *3* (1), 2300103. DOI https://doi.org/10.1002/adsr.202300103.
106. So, F. T.-K.; Hariki, N.; Nemoto, M., et al., *APL Materials* **2024,** *12* (5). DOI 10.1063/5.0201154.
107. Liu, C. F.; Leong, W. H.; Xia, K., et al., *Natl Sci Rev* **2021,** *8* (5), nwaa194. DOI 10.1093/nsr/nwaa194.
108. Bradac, C.; Gao, W.; Forneris, J.; Trusheim, M. E.; Aharonovich, I., *Nature communications* **2019,** *10* (1), 5625. DOI 10.1038/s41467-019-13332-w.
109. Fujiwara, M.; Uchida, G.; Ohki, I., et al., *Carbon* **2022,** *198*, 57–62. DOI https://doi.org/10.1016/j.carbon.2022.06.076.
110. Fedotov, I. V.; Solotenkov, M. A.; Pochechuev, M. S., et al., *ACS Photonics* **2020,** *7* (12), 3353–3360. DOI 10.1021/acsphotonics.0c00706.
111. Plakhotnik, T.; Gruber, D., *Physical Chemistry Chemical Physics* **2010,** *12* (33), 9751–9756. DOI 10.1039/C001132K.
112. Tran, T. T.; Regan, B.; Ekimov, E. A., et al., *Science Advances* **2019,** *5* (5), eaav9180. DOI doi:10.1126/sciadv.aav9180.
113. Alkahtani, M.; Cojocaru, I.; Liu, X., et al., *Applied Physics Letters* **2018,** *112* (24). DOI 10.1063/1.5037053.
114. de Vries, M. O.; del Rosal, B.; Messalea, K. A., et al., *Advanced Sensor Research* **2024,** *3* (1), 2300086. DOI https://doi.org/10.1002/adsr.202300086.
115. Dharmasiri, A.; Vincent, C.; Rajib, T. I., et al., *Optics Letters* **2025,** *50* (3), 968–971. DOI 10.1364/OL.544091.

116. Solotenkov, M. A.; Maltsev, D. I.; Fedotov, A. B., et al., *Optics Express* **2025,** *33* (25), 52304–52320. DOI 10.1364/OE.581799.
117. Li, M.; Zhang, Q.; Kong, X., et al., *Advanced Quantum Technologies* **2024,** *7* (3), 2300293. DOI https://doi.org/10.1002/qute.202300293.
118. Plakhotnik, T.; Doherty, M. W.; Cole, J. H.; Chapman, R.; Manson, N. B., *Nano Letters* **2014,** *14* (9), 4989–4996. DOI 10.1021/nl501841d.
119. Guo, J.; Liu, X.; Wang, Z.; Zhang, N.; Liu, B., *Advanced Quantum Technologies n/a* (n/a), e00215. DOI https://doi.org/10.1002/qute.202500215.
120. Plakhotnik, T.; Aman, H.; Chang, H. C., *Nanotechnology* **2015,** *26* (24), 245501. DOI 10.1088/0957-4484/26/24/245501.
121. Bommidi, D. K.; Pickel, A. D., *Applied Physics Letters* **2021,** *119* (25). DOI 10.1063/5.0072357.
122. Blakley, S.; Liu, X.; Fedotov, I., et al., *ACS Photonics* **2019,** *6* (7), 1690–1693. DOI 10.1021/acsphotonics.9b00206.
123. Dharmasiri, A.; Vincent, C.; Rajib, T. I., et al., *Applied Physics Letters* **2024,** *125* (9). DOI 10.1063/5.0207531.
124. Fan, J.-W.; Cojocaru, I.; Becker, J., et al., *ACS Photonics* **2018,** *5* (3), 765–770. DOI 10.1021/acsphotonics.7b01465.
125. Chen, Y.; White, S.; Ekimov, E. A., et al., *ACS Photonics* **2023,** *10*, 2481–2487. DOI 10.1021/acsphotonics.2c01622.
126. Zhang, J.; He, H.; Zhang, T., et al., *The Journal of Physical Chemistry C* **2023,** *127* (6), 3013–3019. DOI 10.1021/acs.jpcc.2c06966.
127. Choi, S.; Agafonov, V. N.; Davydov, V. A.; Plakhotnik, T., *ACS Photonics* **2019,** *6* (6), 1387–1392. DOI 10.1021/acsphotonics.9b00468.
128. Nguyen, C. T.; Evans, R. E.; Sipahigil, A., et al., *Applied Physics Letters* **2018,** *112* (20), 203102. DOI 10.1063/1.5029904.
129. Hossain, M. S.; Bacaoco, M.; Mai, T. N. A., et al., *ACS Applied Optical Materials* **2024,** *2* (1), 97–107. DOI 10.1021/acsaom.3c00359.
130. Ma, L.; Zhang, J.; Hao, Z., et al., *Advanced Optical Materials* **2024,** *12* (34), 2401232. DOI https://doi.org/10.1002/adom.202401232.
131. Chen, Y.; Li, C.; Yang, T., et al., *ACS Nano* **2023,** *17* (3), 2725–2736. DOI 10.1021/acsnano.2c10974.
132. Golubewa, L.; Padrez, Y.; Malykhin, S., et al., *Advanced Optical Materials* **2022,** *10* (15), 2200631. DOI https://doi.org/10.1002/adom.202200631.
133. Kurochkin, N. S.; Sychev, V. V.; Gritsienko, A. V.; Bi, D., *physica status solidi (RRL) – Rapid Research Letters* **2024,** *18* (1), 2300277. DOI https://doi.org/10.1002/pssr.202300277.
134. Rahman, A. T. M. A.; Frangeskou, A. C.; Kim, M. S., et al., *Scientific Reports* **2016,** *6* (1), 21633. DOI 10.1038/srep21633.
135. Hossain, M. S.; Xu, J.; Mai, T. N. A., et al., *Diamond and Related Materials* **2026,** *163*, 113411. DOI https://doi.org/10.1016/j.diamond.2026.113411.
136. García de Arquer, F. P.; Talapin, D. V.; Klimov, V. I., et al., *Science* **2021,** *373* (6555), eaaz8541. DOI doi:10.1126/science.aaz8541.
137. Resch-Genger, U.; Grabolle, M.; Cavaliere-Jaricot, S.; Nitschke, R.; Nann, T., *Nature Methods* **2008,** *5* (9), 763–775. DOI 10.1038/nmeth.1248.
138. Lee, W. N.; Lai, L.-H.; Tu, Y.-Q., et al., *Materials Today Physics* **2024,** *41*, 101339. DOI https://doi.org/10.1016/j.mtphys.2024.101339.
139. Kalytchuk, S.; Zhovtiuk, O.; Kershaw, S. V.; Zbořil, R.; Rogach, A. L., *Small* **2016,** *12* (4), 466–476. DOI https://doi.org/10.1002/smll.201501984.
140. Medintz, I. L.; Uyeda, H. T.; Goldman, E. R.; Mattoussi, H., *Nature Materials* **2005,** *4* (6), 435–446. DOI 10.1038/nmat1390.

141. Kumbhakar, P.; Roy Karmakar, A.; Das, G. P., et al., *Nanoscale* **2021,** *13* (5), 2946–2954. DOI 10.1039/D0NR07874C.
142. Vlaskin, V. A.; Janssen, N.; van Rijssel, J.; Beaulac, R.; Gamelin, D. R., *Nano Letters* **2010,** *10* (9), 3670–3674. DOI 10.1021/nl102135k.
143. Kato, Y. S.; Shimazaki, Y.; Chuma, S., et al., *Nano Letters* **2025,** *25* (14), 5688–5696. DOI 10.1021/acs.nanolett.4c06642.
144. Duda, M.; Joshi, P.; Borodziuk, A., et al., *Acs Applied Materials & Interfaces* **2024,** *16* (44), 60008–60017. DOI 10.1021/acsami.4c14541.
145. Zhao, Y.; Riemersma, C.; Pietra, F., et al., *ACS Nano* **2012,** *6* (10), 9058–9067. DOI 10.1021/nn303217q.
146. Zhou, J.; Liu, Z.; Li, F., *Chemical Society Reviews* **2012,** *41* (3), 1323–1349. DOI 10.1039/C1CS15187H.
147. Di, X.; Wang, D.; Zhou, J., et al., *Nano Letters* **2021,** *21* (4), 1651–1658. DOI 10.1021/acs.nanolett.0c04281.
148. Di, X.; Wang, D.; Su, Q. P., et al., *Proceedings of the National Academy of Sciences* **2022,** *119* (45), e2207402119. DOI doi:10.1073/pnas.2207402119.
149. Fischer, L. H.; Harms, G. S.; Wolfbeis, O. S., *Angewandte Chemie International Edition* **2011,** *50* (20), 4546–4551. DOI https://doi.org/10.1002/anie.201006835.
150. Yan, L.; Huang, J.; An, Z.; Zhang, Q.; Zhou, B., *Nano Lett* **2022,** *22* (17), 7042–7048. DOI 10.1021/acs.nanolett.2c01931.
151. Sedlmeier, A.; Achatz, D. E.; Fischer, L. H.; Gorris, H. H.; Wolfbeis, O. S., *Nanoscale* **2012,** *4* (22), 7090–7096. DOI 10.1039/C2NR32314A.
152. Rodrigues, E. M.; Gálico, D. A.; Lemes, M. A., et al., *New Journal of Chemistry* **2018,** *42* (16), 13393–13405. DOI 10.1039/C8NJ02471E.
153. Xu, M.; Chen, D.; Huang, P., et al., *Journal of Materials Chemistry C* **2016,** *4* (27), 6516–6524. DOI 10.1039/C6TC02218A.
154. Huang, G.; Wu, X.; Zhan, S.; Liu, Y., *Journal of Materials Chemistry C* **2022,** *10* (13), 5190–5199. DOI 10.1039/D1TC05838J.
155. Wang, Y. F.; Liu, G. Y.; Sun, L. D., et al., *ACS Nano* **2013,** *7* (8), 7200–6. DOI 10.1021/nn402601d.
156. Savchuk, O.; Carvajal, J.; Brites, C., et al., *Nanoscale* **2018,** *10*. DOI 10.1039/C7NR08758F.
157. Green, K.; Huang, K.; Pan, H.; Han, G.; Lim, S. F., *Frontiers in Chemistry* **2018,** *Volume 6 - 2018*. DOI 10.3389/fchem.2018.00416.
158. Yao, L.; Li, Y.; Xu, D., et al., *New Journal of Chemistry* **2019,** *43* (9), 3848–3855. DOI 10.1039/C8NJ06385K.
159. Li, L.; Guo, C.; Jiang, S.; Agrawal, D. K.; Li, T., *RSC Advances* **2014,** *4* (13), 6391–6396. DOI 10.1039/C3RA47264G.
160. Liu, Y.; Duan, X.; Shin, H. J., et al., *Nature* **2021,** *591* (7848), 43–53. DOI 10.1038/s41586-021-03339-z.
161. Taur, Y.; Wann, C. H.; Frank, D. J. In *25 nm CMOS design considerations*, International Electron Devices Meeting 1998. Technical Digest (Cat. No.98CH36217), 6–9 Dec. 1998; **1998**; pp 789–792. DOI 10.1109/IEDM.1998.746474.
162. Saïdi, E.; Babinet, N.; Lalouat, L., et al., *Small* **2011,** *7* (2), 259–264. DOI https://doi.org/10.1002/smll.201001476.
163. Löw, P.; Kim, B.; Takama, N.; Bergaud, C., *Small* **2008,** *4* (7), 908–914. DOI 10.1002/smll.200700581.
164. Hui, Y. Y.; Chen, O. Y.; Azuma, T., et al., *The Journal of Physical Chemistry C* **2019,** *123* (24), 15366–15374. DOI 10.1021/acs.jpcc.9b04496.
165. van Swieten, T. P.; van Omme, T.; van den Heuvel, D. J., et al., *ACS Applied Nano Materials* **2021,** *4* (4), 4208–4215. DOI 10.1021/acsanm.1c00657.

166. Chen, Y.; Tran, T. N.; Duong, N. M. H., et al., *Acs Applied Materials & Interfaces* **2020,** *12* (22), 25464–25470. DOI 10.1021/acsami.0c05735.
167. Öner, B.; Pomeroy, J. W.; Kuball, M., *ACS Applied Electronic Materials* **2020,** *2* (1), 93–102. DOI 10.1021/acsaelm.9b00575.
168. Chen, R.; Shi, B.; Song, K., et al., *Advanced Materials* **2025,** *37* (43), e15604. DOI https://doi.org/10.1002/adma.202415604.
169. Benninger, R. K. P.; Koç, Y.; Hofmann, O., et al., *Analytical Chemistry* **2006,** *78* (7), 2272–2278. DOI 10.1021/ac051990f.
170. Häußler, A. J.; Heller, P.; McGuinness, L. P.; Naydenov, B.; Jelezko, F., *Opt Express* **2014,** *22* (24), 29986–95. DOI 10.1364/oe.22.029986.
171. Badieirostami, M.; Lew, M. D.; Thompson, M. A.; Moerner, W. E., *Appl Phys Lett* **2010,** *97* (16), 161103. DOI 10.1063/1.3499652.
172. Yang, J.; Li, B. Q.; Li, R.; Mei, X., *Nanoscale* **2019,** *11* (5), 2249–2263. DOI 10.1039/C8NR09096C.
173. Ming, L.; Romelli, A.; Lifante, J., et al., *Nature communications* **2025,** *16* (1), 6429. DOI 10.1038/s41467-025-59681-7.
174. Li, D.; Jia, M.; Jia, T.; Chen, G., *Advanced Materials* **2024,** *36* (11), 2309452. DOI https://doi.org/10.1002/adma.202309452.
175. Elliott, A. D., *Current Protocols in Cytometry* **2020,** *92* (1), e68. DOI https://doi.org/10.1002/cpcy.68.
176. Yu, W.; Deschaume, O.; Dedroog, L., et al., *Advanced Functional Materials* **2022,** *32* (5), 2108234. DOI https://doi.org/10.1002/adfm.202108234.
177. Yukawa, H.; Fujiwara, M.; Kobayashi, K., et al., *Nanoscale Advances* **2020,** *2* (5), 1859–1868. DOI 10.1039/D0NA00146E.
178. Choi, J.; Zhou, H.; Landig, R., et al., *Proc Natl Acad Sci U S A* **2020,** *117* (26), 14636–14641. DOI 10.1073/pnas.1922730117.
179. Kushimoto, S.; Gando, S.; Saitoh, D., et al., *Crit Care* **2013,** *17* (6), R271. DOI 10.1186/cc13106.
180. Gerasimova, E.; Cherednikova, A.; Feoktistova, V., et al., *Laser & Photonics Reviews* **2025,** *19* (22), e00691. DOI https://doi.org/10.1002/lpor.202500691.
181. Santos, H. D. A.; Ximendes, E. C.; Iglesias-de la Cruz, M. d. C., et al., *Advanced Functional Materials* **2018,** *28* (43), 1803924. DOI https://doi.org/10.1002/adfm.201803924.
182. Li, S.-J.; Li, F.; Kong, N.; Liu, J.-R.; Zhu, X., *Advanced Healthcare Materials* **2023,** *12* (31), 2302276. DOI https://doi.org/10.1002/adhm.202302276.
183. Carrasco, E.; del Rosal, B.; Sanz-Rodríguez, F., et al., *Advanced Functional Materials* **2015,** *25* (4), 615–626. DOI https://doi.org/10.1002/adfm.201403653.
184. Xu, M.; Zou, X.; Su, Q., et al., *Nature communications* **2018,** *9* (1), 2698. DOI 10.1038/s41467-018-05160-1.
185. Ximendes, E. C.; Rocha, U.; del Rosal, B., et al., *Advanced Healthcare Materials* **2017,** *6* (4), 1601195. DOI https://doi.org/10.1002/adhm.201601195.
186. Wu, Y.; Li, F.; Wu, Y., et al., *Nature communications* **2024,** *15* (1), 2341. DOI 10.1038/s41467-024-46727-5.
187. Zhang, Y.; Hong, G.; Zhang, Y., et al., *ACS Nano* **2012,** *6* (5), 3695–3702. DOI 10.1021/nn301218z.
188. Yen, A., *Journal of Micro/Nanopatterning, Materials, and Metrology* **2021,** *20* (1), 010501.
189. Pearce, A. K.; O'Reilly, R. K., *Bioconjug Chem* **2019,** *30* (9), 2300–2311. DOI 10.1021/acs.bioconjchem.9b00456.
190. Zhou, J.; Wen, S.; Liao, J., et al., *Nature Photonics* **2018,** *12* (3), 154–158. DOI 10.1038/s41566-018-0108-5.

191. van Swieten, T. P.; van Omme, T.; van den Heuvel, D. J., et al., *ACS Appl Nano Mater* **2021,** *4* (4), 4208–4215. DOI 10.1021/acsanm.1c00657.
192. Stone, D. G.; Chen, Y.; Ekimov, E. A.; Tran, T. T.; Bradac, C., *ACS Applied Optical Materials* **2023,** *1* (4), 898–905. DOI 10.1021/acsaom.3c00059.
193. Maturi, F. E.; Brites, C. D. S.; Ximendes, E. C., et al., *Laser & Photonics Reviews* **2021,** *15* (11), 2100301. DOI https://doi.org/10.1002/lpor.202100301.
194. Hossain, M. S.; Stone, D. G.; Landry, G., et al., *Materials Today Communications* **2026,** *52*, 115196. DOI https://doi.org/10.1016/j.mtcomm.2026.115196.
195. Xiao, Y.; Wan, C.; Shahsafi, A.; Salman, J.; Kats, M. A., *ACS Photonics* **2020,** *7* (4), 853–860. DOI 10.1021/acsphotonics.9b01588.
196. Marciniak, L.; Szymczak, M.; Woźny, P.; Runowski, M., *Advanced Optical Materials* **2025,** *13* (31), e00914. DOI https://doi.org/10.1002/adom.202500914.
197. Goderski, S.; Runowski, M.; Woźny, P.; Lavín, V.; Lis, S., *Acs Applied Materials & Interfaces* **2020,** *12* (36), 40475–40485. DOI 10.1021/acsami.0c09882.